\documentclass[twocolumn,british,prl,superscriptaddress]{revtex4-2}
\usepackage[T1]{fontenc}
\usepackage[utf8]{inputenc}
\usepackage{amsmath}
\usepackage{amssymb}
\usepackage{graphicx}
\usepackage{babel}
\begin{document}
\title{A microscopic approach to nonlinear theory of spin-charge separation}
\author{Oleksandr Tsyplyatyev}
\affiliation{Institut f\"ur Theoretische Physik, Universit\"at Frankfurt, Max-von-Laue-Stra{\ss}e 1, 60438 Frankfurt, Germany}
\author{Yiqing Jin}
\affiliation{Department of Physics, Cavendish Laboratory, University of Cambridge, Cambridge, CB3 0US, UK}
\author{Mar\'ia Moreno}
\affiliation{Department of Physics, Cavendish Laboratory, University of Cambridge, Cambridge, CB3 0US, UK}
\affiliation{Departamento de F\'isica Aplicada, Universidad de Salamanca, Plaza de la Merced s/n, 37008 Salamanca, Spain}
\author{Wooi Kiat Tan}
\affiliation{Department of Physics, Cavendish Laboratory, University of Cambridge, Cambridge, CB3 0US, UK}
\author{Christopher J.B. Ford}
\affiliation{Department of Physics, Cavendish Laboratory, University of Cambridge, Cambridge, CB3 0US, UK}
\begin{abstract}
The fate of spin-charge separation beyond the low energy remains elusive up to now. Here we develop a microscopic theory of the correlation functions using the strong coupling expansion of the Hubbard model and demonstrate its validity down to the experimentally relevant $r_{\rm s}>1$. Evaluating the spectral function, we show the general stability of the nonlinear spin-charge modes in whole energy band and investigate all the nonlinear features systematically. We confirm the general prediction experimentally in semiconductor quantum wires. Furthermore, we observe a signal consistent with a continuum of the nonlinear excitations and with a final spectral density around the $3 k_{\rm F}$ point, indicating the robustness of the Hubbard model predictions for a finite range interaction. 
\end{abstract}
\maketitle

Interactions restructure completely the many-body spectrum of electrons in one dimension (1D), resulting in the formation of the Luttinger liquid \cite{Tomonaga50,Luttinger63} instead of the Fermi liquid already at low energy. Such a dramatic change manifests itself in the appearance of the pseudo-gap and in the separation of the spin and charge excitations \cite{Schoenhammer92,Voit93}, both of which were confirmed experimentally in a variety of systems \cite{Kim96,Zwick98,Yao99,Auslaender02,Venkataraman06,Jompol09,Hashisaka17,Salomon19,Vijayan20,Weldeyesys25,BouchouleReview25}. The fate of these effects beyond the linear regime remains unknown, with attempts via field theory blocked by many divergences \cite{Samokhin98} or being inconclusive \cite{Schmidt10,Veness16,Carmelo17,Mestyan19,Nozawa20,Luo24,Patu25}. On the other hand, the microscopic approach via the Hubbard model has been partially successful, as the exact Lieb-Wu solution \cite{LiebWu68} allowed the calculation of the full spectrum \cite{Schulz90,Schulz95,Vianez21}, but a direct attempt to bring the algebraic method \cite{Gardner67,Lax68}, developed for the correlation functions of spin chains \cite{Fadeev79,Drinfeld87}, to the Hubbard model still failed \cite{Shastry86}.

Here we choose a different path of constructing the $t/U$ expansion for the correlation functions starting from the $U=\infty$ point, where the Lieb-Wu wave functions factorise into the spin and charge sectors \cite{Ogata90} allowing to use the algebraic method with only some adjustments \cite{Tsyplyatyev22}. Evaluating the occupation numbers, we show the validity of such an expansion down to the interaction strengths $r_{\rm s}>1$, and from the spectral function we find the general stability of the spin-charge-separated modes in the whole energy band and systematically investigate the nonlinear features. Testing these predictions for the realistic screened-Coulomb interaction, we measure semiconductor quantum wires as in \cite{Jompol09} using the magnetotransport spectroscopy technique \cite{Hayden91,Wang_resonant_1994,Kardinal96} and find a signal consistent with a broad continuum of the nonlinear excitations around the charge mode in the particle sector as well as a finite spectral density around the $3k_{\rm F}$ point, indicating experimentally the reliability of the Hubbard-model predictions for a finite-range interaction. This shows spin-charge splitting of the whole energy band, presenting a novel mechanism for band-structure engineering based solely on interactions in a simple crystal.

We analyse the 1D Hubbard model describing electrons with short-range interaction,
\begin{equation}
H=-t\sum_{j\alpha}\left(c_{j\alpha}^{\dagger}c_{j+1,\alpha}+c_{j\alpha}^{\dagger}c_{j-1,\alpha}\right)+U\sum_{j}n_{j\uparrow}n_{j\downarrow},\label{eq:H}
\end{equation}
where $c_{j\alpha}$ are the Fermi operators at site $j$ for the \mbox{spin-1/2} index $\alpha=\uparrow\,\textrm{or}\,\downarrow$,
$n_{j\alpha}=c_{j\alpha}^{\dagger}c_{j\alpha}$
is the local-density operator for the spin species $\alpha$, $t$
is the hopping amplitude, $U>0$ is the repulsive on-site interaction
energy, and we consider the periodic boundary condition, $c_{j+L}=c_{j}$,
for a chain of length $L$. This model was diagonalised exactly in the
$N$-particle sector by Lieb and Wu \citep{LiebWu68} via the solutions
of a set of nonlinear equations for $N$ charge quasimomenta $\mathbf{k}=\left(k_{1},\dots,k_{N}\right)$
and $M$ spin quasimomenta $\mathbf{q}=\left(q_{1},\dots,q_{M}\right)$,
which give the corresponding eigenenergies ($E=t\sum_{j}(k_{j})^{2}/2$), total momenta ($P=\sum_{j}k_{j}$), and eigenstates as the Lieb-Wu wave functions $|\Psi\rangle$ \citep{SM}. Here, we consider only the low-particle densities $N/L\ll1$ in the thermodynamic limit $N,L\gg1$.

In the infinite-interaction limit, $U=\infty$, the Lieb-Wu wave functions
factorise, $\left|\Psi^{0}\right\rangle =\left|\Psi_{\rm c}^{0}\right\rangle \otimes\left|\Psi_{\rm s}^{0}\right\rangle$,
into product of a Slater determinant $\left|\Psi_{\rm c}^{0}\right\rangle $ in the
charge and a Bethe wave function $\left|\Psi_{\rm s}^{0}\right\rangle $
in the spin sector \citep{Ogata90}. The nonlinear equations for the
spins also separate out into  independent Bethe equations, $Nq_{m}^{0}-2\sum_{l\neq m}\varphi_{ml}=2\pi J_{m}$,
where $e^{i2\varphi_{lm}}=-\big(e^{iq_{l}^{0}+iq_{m}^{0}}+1-2e^{iq_{l}^{0}}\big)/\big(e^{iq_{l}^{0}+iq_{m}^{0}}+1-2e^{iq_{m}^{0}}\big)$
are two-spinon scattering phases, simplifying the charge equations
to single-particle quantisation conditions, $Lk_{j}^{0}-P_{\rm s}=2\pi I_{j}$,
where $P_{\rm s}=\sum_{m}q_{m}^{0}$ is the total spin momentum. For a
large $U/t$ these solutions depart slightly from the $U=\infty$
limit so that such deviations can be linearised in $t/U$ as $\mathbf{k}=\mathbf{k}^{0}+\mathbf{k}^{1}t/U$
and $\mathbf{q}=\mathbf{q}^{0}+\mathbf{q}^{1}t/U$. In turn, linearisation
of the nonlinear equations gives only a single-particle correction
to the charge and a linear set of equations (that is solved via matrix
inversion) for the spin quasimomenta in the first $t/U$-order, 
\begin{align}
\mathbf{k}^{1}= & \mathbf{k}^{0}\frac{4}{L}\sum_{m}\left(\cos q_{m}^{0}-1\right),\label{eq:k1}\\
\mathbf{q}^{1}= & 4P\hat{Q}_{0}^{-1}\left(\mathbf{1}-\cos\mathbf{q}^{0}\right),\label{eq:q1}
\end{align}
where the matrix elements of $\hat{Q}_{0}$ are $Q_{aa}^{0}=N-\sum_{l\neq a}4\left(1-\cos q_{l}^{0}\right)/\big(e^{iq_{l}^{0}}+e^{-iq_{a}^{0}}-2\big)/\big(e^{-iq_{l}^{0}}+e^{iq_{a}^{0}}-2\big)$
and $Q_{ab}^{0}=4\left(1-\cos q_{b}^{0}\right)/\big(e^{iq_{b}^{0}}+e^{-iq_{a}^{0}}-2\big)/\big(e^{-iq_{b}^{0}}+e^{iq_{a}^{0}}-2\big)$
for $a\neq b$. 

Comparison of the zeroth with the first order for the charge quasimomenta
in Eq.\ (\ref{eq:k1}) gives a dimensionless parameter $\gamma=-Uk_{j}^{0}/(k_{j}^{1}t)$
that is independent of $j$. Evaluation of the sum over $m$ for the
ground state in the thermodynamic limit, $\sum_{m}\left(\cos q_{m}^{0}-1\right)/N=-0.69(3)$,
gives \citep{Tsyplyatyev14}
\begin{equation}
\gamma=0.09\lambda_{\rm F}\frac{U}{t},\label{eq:gamma_definition}
\end{equation}
in which the Fermi wavelength $\lambda_{\rm F}=4L/N$ appears as an extra
large factor in addition to $U/t$ in the validity of this expansion.
The same extra factor appears for the spin quasimomenta in Eq.\ (\ref{eq:q1}),
for which a typical $P\lesssim N/L$ and $[\hat{Q}_{0}^{-1}(\mathbf{1}-\cos\mathbf{q}^{0})]_m\simeq1$,
making $1/\gamma$ the emergent small parameter controlling the large-$U$
expansion of the nonlinear eigenvalue problem. The factor $\lambda_{\rm F}$
in Eq.\ (\ref{eq:gamma_definition}) is similar to the generic parameter
of the Coulomb interaction $r_{\rm s}\propto\lambda_{\rm F}$ \citep{SM},
allowing us to interpret $\gamma\simeq r_{\rm s}$ as the microscopic
calculation of the phenomenologically introduced $r_{\rm s}$ in 1D \citep{Vianez21}.

The Lieb-Wu wave function can also be linearised in $t/U$ as $\left|\Psi\right\rangle =\left|\Psi^{0}\right\rangle +t/U\left|\Psi^{1}\right\rangle $.
This translates into the expansion matrix element $\left\langle f|c_{1\alpha}^{\pm}|0\right\rangle =\left\langle f|c_{1\alpha}^{\pm}|0\right\rangle ^{0}+t/U\left\langle f|c_{1\alpha}^{\pm}|0\right\rangle ^{1}$
needed for the correlation functions, \emph{e.g.}, for the Green function
$G_{\alpha}\left(k,E\right)=\sum_{f}[|\langle f|c_{k\alpha}^{+}|0\rangle|^{2}/(E-E_{f}+i\eta)+|\langle f|c_{k\alpha}|0\rangle|^{2}/(E+E_{f}-i\eta)]$,
where $0=\left(\mathbf{k}^{0},\mathbf{q}^{0}\right)$ and $f=\left(\mathbf{k}^{f},\mathbf{q}^{f}\right)$
are the ground state and an excited state, $c_{k}^{\pm}=\sum_{j}c_{j}^{\pm}e^{\pm ikj}/\sqrt{L}$
is the Fourier transform, $\langle f|c_{k\alpha}^{\pm}|0\rangle=\langle f|c_{1\alpha}^{\pm}|0\rangle\delta\left(k\pm P_{f}\right)$
due to the translational symmetry, $\eta$ is an infinitesimally small
real number, and without external magnetic field $G_{\uparrow}\left(k,E\right)=G_{\downarrow}\left(k,E\right)$
so we can consider only $\alpha=\,\uparrow$ here. The leading term
factorises as $\langle f|c_{1\uparrow}|0\rangle^{0}=\langle f|c_{1\uparrow}|0\rangle_{\rm c}^{0}\cdot\langle f|c_{1\uparrow}|0\rangle_{\rm s}^{0}$,
since the wave function factorises in the $U/t=\infty$ limit. The
charge part was evaluated using the first quantisation directly in
\citep{Ogata90} giving, \emph{e.g.},
$\langle f|c_{1\uparrow}|0\rangle_{\rm c}^{0}=L^{-N+1/2}\det\hat{C}_{0}$
with the matrix elements $C_{a1}^{0}=1$ and $C_{ab}^{0}=2\big(k_{a}^{00}-k_{b-1}^{f0}\big)^{-1}\sin\big(\frac{P_{\rm s}^{0}-P_{\rm s}^{f}}{2}\big)$
for $b>1$. The spin part was evaluated using the algebraic Bethe
ansatz to deal with the Bethe wave function in \citep{Tsyplyatyev22}
giving $\langle f|c_{1\uparrow}|0\rangle_{\rm s}^{0}=Z_{0}^{-1}Z_{f}^{-1}\prod_{lm}\big(e^{iq_{l}^{f0}}+e^{-iq_{m}^{00}}-2\big)\prod_{l\neq m}\big(e^{iq_{l}^{f0}}+e^{-iq_{m}^{f0}}-2\big)^{-\frac{1}{2}}\prod_{l\neq m}\big(e^{iq_{l}^{00}}+e^{-iq_{m}^{00}}-2\big)^{-\frac{1}{2}}\det\hat{R}_{00}$
with $R_{Mb}^{00}=e^{ik_{b}^{00}}\prod_{l\neq b}\big(e^{iq_{l}^{00}}+e^{-iq_{b}^{00}}-2\big)/\prod_{l}\big(e^{iq_{l}^{f0}}+e^{-iq_{b}^{00}}-2\big)$ and
$R_{ab}^{00}=\big[e^{iq_{b}^{00}\left(N-1\right)}\prod_{l\neq a}\big(e^{iq_{l}^{f0}+iq_{b}^{00}}+1-2e^{iq_{l}^{f0}}\big)/\big(2e^{iq_{b}^{00}}-e^{iq_{l}^{f0}+iq_{b}^{00}}-1\big)-1\big]/\big(e^{-q_{a}^{f0}}-e^{-iq_{b}^{00}}\big)/\big(e^{q_{a}^{f0}}-e^{-iq_{b}^{00}}-2\big)$
for $a<M$. Here $Z_{0/f}^{2}=\det\hat{Q}_{0}^{0/f}$ are the Gaudin
normalisation factors \citep{Gaudin81}. 

The linear term in the wave function comes from three sources, $|\Psi^{1}\rangle=|\Psi_{\rm s}^{1}\rangle+|\Psi_{\rm c}^{1}\rangle+|\Psi_{sc}^{1}\rangle$,
two are the expansions of $|\Psi^{0}\rangle$ in the linear terms
of quasimomenta in Eqs.~(\ref{eq:k1},\ref{eq:q1}) that do not
break the spin-charge factorisation of $|\Psi^{0}\rangle$ and one
is the mixing term $|\Psi_{sc}^{1}\rangle$. This makes the linear
term of the matrix element also a linear superposition of the same
three contributions $\langle f|c_{1\uparrow}|0\rangle^{1}=\langle f|c_{1\uparrow}|0\rangle_{\rm c}^{1}+\langle f|c_{1\uparrow}|0\rangle_{\rm cs}^{1}+\langle f|c_{1\uparrow}|0\rangle_{\rm s}^{1}$.
The charge part comes from the $t/U$ expansion of the bra and ket
states, $\langle f|c_{1\uparrow}|0\rangle_{\rm c}^{1}=(\langle f_{\rm c}^{0}|c_{1\uparrow}|0_{\rm c}^{1}\rangle+\langle f_{\rm c}^{1}|c_{1\uparrow}|0_{\rm c}^{0}\rangle)\langle f|c_{1\uparrow}|0\rangle_{\rm s}^{0}$.
We evaluate both contributions simultaneously by changing $\mathbf{k}_{0/f}^{0}\rightarrow\mathbf{k}_{0/f}^{0}+g\mathbf{k}_{0/f}^{1}$
in $\langle f|c_{1\uparrow}|0\rangle_{\rm c}^{0}$, repeating the same
calculation as for the $\langle f|c_{1\uparrow}|0\rangle_{\rm c}^{0}$
as in \citep{Ogata90}, and taking a derivative and a limit of the
resulting determinant expression as $\lim_{g\rightarrow0}\partial_{g}\det\big(\hat{C}_{0}+g\hat{C}_{1}\big)=\det\hat{C}_{0}\textrm{Tr}\big(\hat{C}_{0}^{-1}\hat{C}_{1}\big)$
using generic matrix identities. For the whole charge part, we find
that the $U/t=\infty$ matrix element appears as a factor $\langle f|c_{1\uparrow}|0\rangle_{\rm c}^{1}=\langle f|c_{1\uparrow}|0\rangle^{0}T_{\rm c}$
and 
\begin{equation}
T_{\rm c}={\rm Tr}\big(\hat{C}_{0}^{-1}\hat{C}_{1}\big),\label{eq:Tc}
\end{equation}
where the derivatives of the entries of $\hat{C}_{0}$ under the shifts
by $g\mathbf{k}_{0/f}^{1}$ are 
\begin{align}
C_{a1}^{1}= & -i\frac{k_{a}^{01}\big(L-1\big)}{2},\\
C_{ab}^{1}= & 2\Bigg(\frac{L\cos\frac{P_{\rm s}^{0}-P_{\rm s}^{f}}{2}}{2}-\frac{\sin\frac{P_{\rm s}^{0}-P_{\rm s}^{f}}{2}}{k_{a}^{00}-k_{b-1}^{f0}}\Bigg)\frac{k_{a}^{01}-k_{b-1}^{f1}}{k_{a}^{00}-k_{b-1}^{f0}},
\end{align}
 for $b>1$. 

We evaluate the spin part $\left\langle f|c_{k\uparrow}|0\right\rangle _{\rm s}^{1}$
using the same trick with shifting $\mathbf{q}_{0/f}^{0}\rightarrow\mathbf{q}_{0/f}^{0}+g\mathbf{q}_{0/f}^{1}$
in $\left\langle f|c_{1\uparrow}|0\right\rangle _{\rm s}^{0}$, repeating
the same calculation as for the $\langle f|c_{1\uparrow}|0\rangle_{\rm s}^{0}$
in \citep{Tsyplyatyev22}, and taking $\lim_{g\rightarrow0}\partial_{g}$
and obtain analogously $\left\langle f|c_{1\uparrow}|0\right\rangle _{\rm s}^{1}=\left\langle f|c_{1\uparrow}|0\right\rangle ^{0}T_{\rm s}$
with
\begin{multline}
T_{\rm s}=i\sum_{lm}\frac{q_{l}^{f1}e^{iq_{l}^{f0}}-q_{m}^{01}e^{-iq_{m}^{00}}}{e^{iq_{l}^{f0}}+e^{-iq_{m}^{00}}-2}\\
-\frac{i}{2}\sum_{l\neq m}\frac{q_{l}^{01}e^{iq_{l}^{00}}-q_{m}^{01}e^{-iq_{m}^{00}}}{e^{iq_{l}^{00}}+e^{-iq_{m}^{00}}-2}+{\rm Tr}\big(\hat{R}_{00}^{-1}\hat{R}_{01}\big)\\
-\frac{i}{2}\sum_{l\neq m}\frac{q_{l}^{f1}e^{iq_{l}^{f0}}-q_{m}^{f1}e^{-iq_{m}^{f0}}}{e^{iq_{l}^{f0}}+e^{-iq_{m}^{f0}}-2}+{\rm Tr}\big(\hat{R}_{f0}^{-1}\hat{R}_{f1}\big),\label{eq:Ts}
\end{multline}
where the extra sums appear due to a different normalisation of the
Bethe wave functions in the algebraic representation, the second trace
with the matrix $R_{aM}^{f0}=1$ and $R_{ab}^{f0}=\big[e^{iq_{b}^{00}N}\prod_{l\neq a}\big(e^{iq_{b}^{f0}+iq_{l}^{00}}+1-2e^{iq_{l}^{00}}\big)/\big(2e^{iq_{b}^{f0}}-e^{iq_{b}^{f0}+iq_{l}^{00}}-1\big)-1\big]/\big(e^{-q_{b}^{f0}}-e^{-iq_{a}^{00}}\big)/\big(e^{q_{a}^{00}}-e^{-iq_{b}^{f0}}-2\big)$
for $b<M$ is due to extra mathematical complications in applying the
Slavnov's formula \citep{Slavnov89} to linear expansion of the bra
and ket states of the spin matrix element, and the derivatives of the entries of $\hat{R}_{0/f,0}$
shifted by $g\mathbf{q}_{0/f}^{1}$
are $R_{ab}^{0/f,1}=\sum_{j}q_{j}^{0/f,1}\partial_{q_{j}^{0/f,0}}R_{ab}^{0/f,0}$. 

In the mixing part $\langle f|c_{1\uparrow}|0\rangle_{\rm cs}^{1}$, the
spin and charge coordinates mix in a linear way. After summation over
the charge coordinates, the remaining spin dependence has a part proportional
to $\left\langle f|c_{1\uparrow}|0\right\rangle ^{0}$ that gives
$\left\langle f|c_{1\uparrow}|0\right\rangle _{\rm cs}^{1}=\left\langle f|c_{1\uparrow}|0\right\rangle ^{0}T_{\rm cs}$, with
\begin{multline}
T_{\rm cs}=P_{f}\Big[4\frac{\sum_{m}\cos q_{m}^{f0}-\sum_{m}\cos q_{m}^{00}+\cos\left(P_{\rm s}^{f}-P_{\rm s}^{0}\right)}{\tan\frac{P_{\rm s}^{f}-P_{\rm s}^{0}}{2}}\\
+2i\frac{\sum_{m}e^{iq_{m}^{00}}-\sum_{m}e^{-iq_{m}^{f0}}}{N}+2ie^{i\left(P_{\rm s}^{f}-P_{\rm s}^{0}\right)}\Big].\label{eq:Tcs}
\end{multline}
The other part of the spin dependence has the same structure as $\left\langle f|c_{1\uparrow}|0\right\rangle_{\rm s}^{1}$
but with $\mathbf{q}_{0/f}^{1}=4P_{f}(\cos \mathbf{q}_{0/f}^{0}-\mathbf{1})/N$
instead of the linear term of the spin quasimomenta in Eq.\ (\ref{eq:q1}),
giving the same result as in Eq.\ (\ref{eq:Ts}) but with different
values of $\mathbf{q}_{0/f}^{1}$. Therefore, in the linear term of
the whole matrix element the zeroth-order matrix element appears as
a factor, $\left\langle f|c_{1\uparrow}|0\right\rangle ^{1}=\left\langle f|c_{1\uparrow}|0\right\rangle ^{0}T$
with $T=T_{\rm c}+T_{\rm s}+T_{\rm cs}$, in which $T_{\rm c}$ and $T_{\rm cs}$ are
given by Eqs.\ (\ref{eq:Tc}, \ref{eq:Tcs}) and $T_{\rm s}$ is given by
Eq.\ (\ref{eq:Ts}), where the second mixing part is added to quasimomenta
in Eq.\ (\ref{eq:q1}) as $\mathbf{q}_{0/f}^{1}\rightarrow\mathbf{q}_{0/f}^{1}+4P_{f}(\cos\mathbf{q}_{0/f}^{0}-\mathbf{1})/N$
since this contribution to the matrix element is linear in $\mathbf{q}_{0/f}^{1}$.

The normalisation of the Lieb-Wu wave functions in this work was chosen
to be unity in the $U=\infty$ limit, $\langle\Psi^{0}|\Psi^{0}\rangle=1$,
in the same way as in \citep{Tsyplyatyev22}. To the linear order that we calculate here,
the normalisation changes as $\langle\Psi|\Psi\rangle=1+\delta Zt/U$,
which we evaluate using the same linear expansion of the Lieb-Wu functions
and the same methods as for the matrix element above. We find that
the charge part does not contribute and the spin together with the
mixing part give
\begin{equation}
\delta Z=2{\rm Re}{\rm Tr}\big(\hat{Q}_{0}^{-1}\hat{Q}_{1}\big)-\frac{4N}{L}\sum_{m}\big(\cos q_{m}^{0}-1\big),\label{eq:dZ}
\end{equation}
where $\hat{Q}_1$ are the derivatives of entries of $\hat{Q}_{0}$ shifted by $g\mathbf{q}^{1}$, which are presented in \cite{SM}.
Combining this result with the linear expansion above, we
find the modulus squared needed for the observables as $|\langle f|c_{1\uparrow}|0\rangle|^{2}=|\langle f|c_{1\uparrow}|0\rangle^{0}|^{2}\big[1+\big(2\textrm{Re}T-\delta Z_{f}-\delta Z_{0}\big)t/U\big]$.
The full details on the derivations of Eqs.\ (\ref{eq:Tc}--\ref{eq:dZ})
are in \citep{SM}. The expressions for the matrix element $\langle f|c_{1\uparrow}^{+}|0\rangle$
in the linear order are the same as in Eqs.\ (\ref{eq:Tc}--\ref{eq:dZ}),
in which the quasimomenta are swapped as $\mathbf{k}^{f},\mathbf{q}^{f}\leftrightarrow\mathbf{k}^{0},\mathbf{q}^{0}$
and the particle and spin quantum numbers are increased by one, $N\rightarrow N+1$
and $M\rightarrow M+1$ as in the zeroth order \cite{Tsyplyatyev22}.

\begin{figure}
\centering
\includegraphics[width=1\columnwidth]{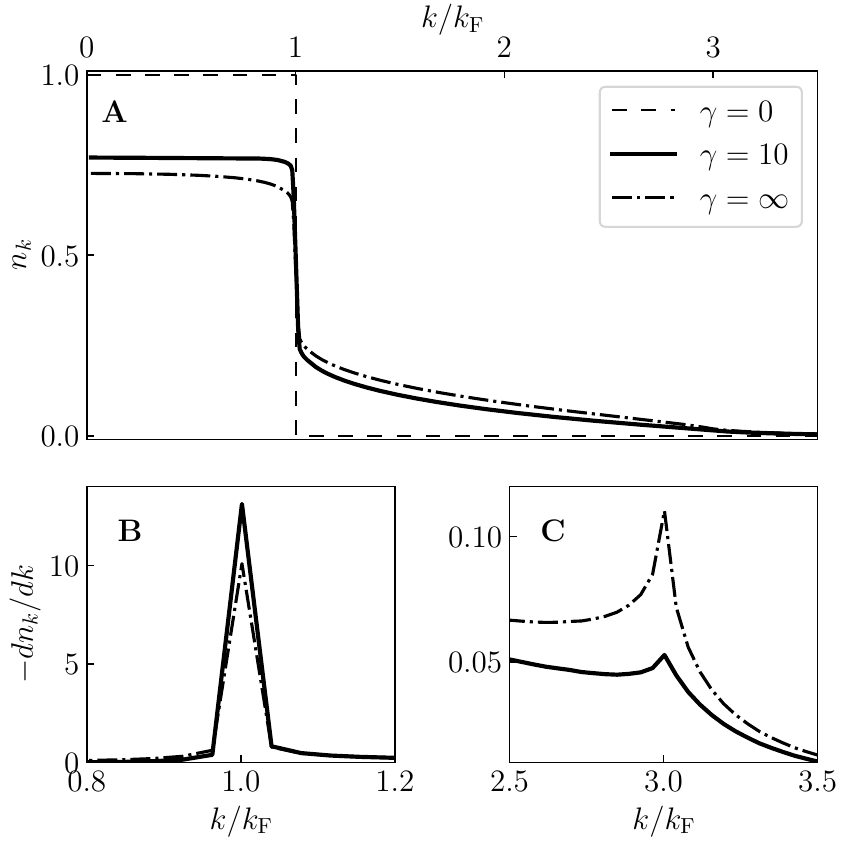}\caption{\textbf{A} Occupation numbers $n_{k}$ evaluated using Eqs.~(\ref{eq:Tc}--\ref{eq:dZ})
for $N=100$ particles. The dash-dotted line is the infinite-interaction
limit $\gamma=\infty$, the solid line is at finite interaction $\gamma=10$,
and the dashed line is the Fermi step of the free-particle limit $\gamma=0$.
\textbf{B} and \textbf{C} is the derivative of $n_{k}$ in \textbf{A} with respect
to $k$, exhibiting singularities around the $k_{\rm F}$ and $3k_{\rm F}$ points, respectively.}\label{fig:nk}
\end{figure}
Analysing correlation functions at a finite $U$, we start from the
occupation numbers $n_{k}=\sum_{f}|\langle f|c_{k\uparrow}|0\rangle|^{2}$.
Numerical evaluation of the sum over $f$ using the leading two levels,
$l\leq1$, of the hierarchy of modes \cite{Tsyplyatyev15,*Tsyplyatyev16}
is presented in Fig.\ \ref{fig:nk} as a solid line. Above $k_{\rm F}$,
we find that a part of the spectral power, which was redistributed
there at $U=\infty$ (dash-dotted line) to form the second Fermi point
at $3k_{\rm F}$ \citep{Tsyplyatyev22}, moves back below $k_{\rm F}$,
recovering the Fermi function at $U=0$ (the dashed line). The $3k_{\rm F}$
Fermi point itself (which appears as a divergence in the first derivative
$\partial_{k}n_{k}$ for Luttinger liquids \citep{Luttinger60,Haldane93})
remains stable away from the $U=\infty$ limit, see Fig.\ \ref{fig:nk}C.
Further, we analyse the part of the spectral weight above $k_{\rm F}$
per particle, $\mathcal{I}=2\int_{k_{\rm F}}^{\infty}{\rm d}kn_{k}/N$, to assess the validity of the $t/U$ expansion for the correlation functions. Evaluating the integral over $k$ numerically for different $N$ and
$L$, we find an additional $N/L$ factor in the linear $t/U$-term
that can be absorbed into $\gamma$ defined in Eq.\ (\ref{eq:gamma_definition}),
and obtain $\mathcal{I}=0.25+0.05/\gamma$. This demonstrates that the expansion for the correlation functions is controlled by the same $\gamma$ as the expansion of the eigenvalues in Eqs.~(\ref{eq:k1}, \ref{eq:q1}). 

\begin{figure}
\centering
\includegraphics[width=1\columnwidth]{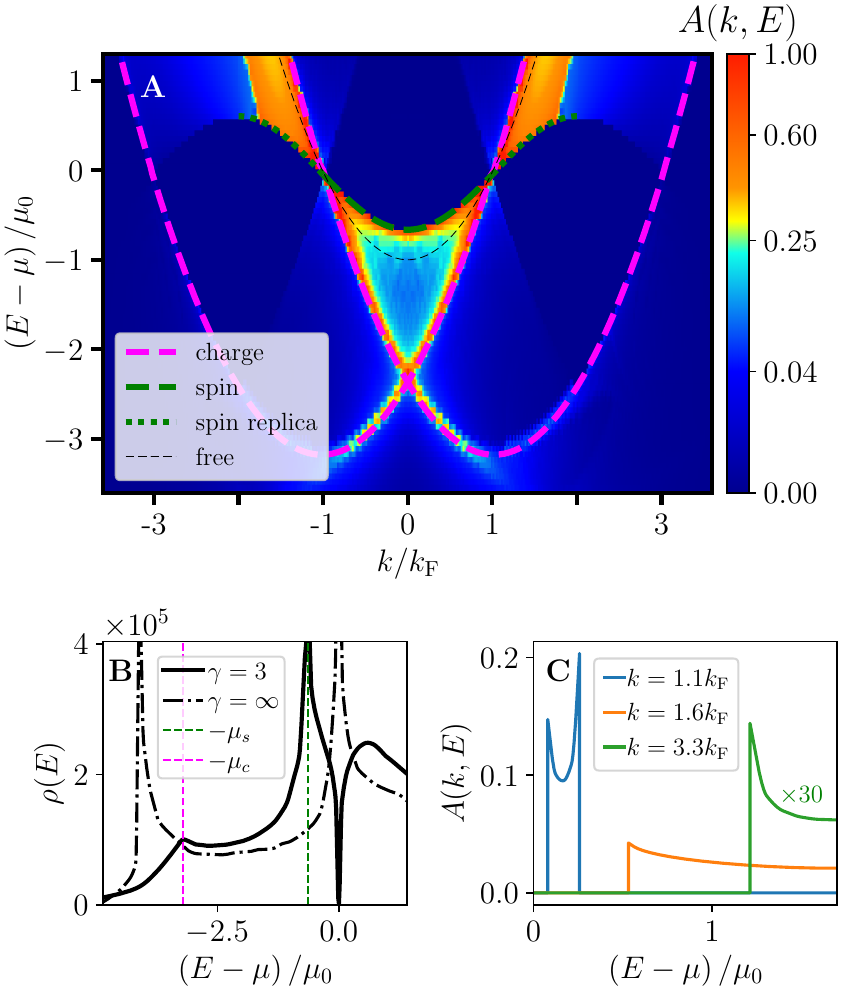}\caption{\textbf{A} Spectral function $A\left(k,E\right)$ evaluated for $N=200$
particles and a finite interaction strength $\gamma=3$. The magenta
and green dashed lines are the dispersions of the pure charge and
spin modes evaluated using the expansion in Eqs.\ (\ref{eq:k1},\ref{eq:q1}).
The green dotted line is the replica of the spin mode in the
hole sector in the $E>\mu$ and $k>k_{\rm F}$ region. The black
dashed line is the free-particle dispersion for $\gamma=0$. \textbf{B}
Density of states $\rho\left(E\right)$ for $\gamma=3$ (solid black line) and $\gamma=\infty$ (dash-dotted
black line). The magenta and green dashed lines mark the charge ($-\mu_{\rm c}$)
and the spin ($-\mu_{\rm s}$) chemical potentials for $\gamma=3$ obtained
as the minimum energy of the charge and spin dispersions w.r.t.\ the
electron chemical potential $\mu$ in \textbf{A}. \textbf{C} Constant-momentum
cuts of $A\left(k,E\right)$ in \textbf{A} around the Fermi point
at $k=1.1k_{\mathbf{F}}$, of the nonlinear extension of the spin
mode in the particle sector at $k=1.6k_{\rm F}$, and of the nonlinear
charge mode above the $3k_{\rm F}$ point at $k=3.3k_{\rm F}$.}\label{fig:spectral_function}
\end{figure}
Now, using the result in Eqs.\ (\ref{eq:Tc}--\ref{eq:dZ}), we evaluate
another observable -- the spectral function $A(k,E)=\sum_{f}|\langle f|c_{k\uparrow}^{+}|0\rangle|^{2}\delta(E-E_{f}+E_{0})+\sum_{f}|\langle f|c_{k\uparrow}|0\rangle|^{2}\delta(E+E_{f}-E_{0})$
-- in Fig.\ \ref{fig:spectral_function}A. Around the $\pm k_{\rm F}$
points, there are two singular peaks with different velocities and
a ``shadow band'' described by the linear Tomonaga-Luttinger theory
\citep{Schoenhammer92,Voit93}. Away from them, both peaks generally
remain stable, see the dashed green and magenta lines,
showing the splitting of the whole single-particle band (the black
dashed line) into two by interactions. On a more detailed level, the
nonlinear parts of these modes are asymmetric w.r.t.\ the electronic
chemical potential $\mu$. In the hole sector $E<\mu$, the whole
spin mode 
remains stable
but the charge mode becomes unstable at the bottom of its dispersion.
This instability is also apparent in the density of states $\rho\left(E\right)=L\int {\rm d}kA(k,E)$,
see the full black line in Fig.\ \ref{fig:spectral_function}B. The
van Hove singularity at the spin chemical potential $-\mu_{\rm s}$ (defined
as the distance from $\mu$ to the bottom of the green dispersion
in Fig.\ \ref{fig:spectral_function}A) remains stable but the van
Hove singularity of the charge mode at $-\mu_{\rm c}$ disappears at a
finite $U$. The latter remains a singularity only in the $U=\infty$
limit, see the dash-dotted line in Fig.\ \ref{fig:spectral_function}B,
in which the spin singularity also contributes to the low-energy behaviour around $E=\mu$
since the spin-mode dispersion is completely flat $\mu_{\rm s}=0$ \cite{Schulz95,Penc96,Tsyplyatyev22}. At a finite $U$, $\mu_{\rm s}$ becomes finite, revealing
the power-law vanishing of $\rho\left(E\right)$,
well-known from the linear theory \citep{Giamarchi_book}.

In the particle sector $E>\mu$, the whole charge mode remains stable
but the spin mode becomes a weaker singularity, only a jump instead
of a singular peak, see the orange cut of $A\left(k,E\right)$ in
Fig.~\ref{fig:spectral_function}C. The states forming the latter
mode always have a pair of degenerate spin quasimomenta making $\langle f|c_{1\uparrow}|0\rangle_{\rm s}^{0}=0$
in the $l=0$ level of the hierarchy. However, the states from the
continuum of the $l=1$ level do not have such a degeneracy and their
squeezing from a wide $k_{\rm F}<k<3k_{\rm F}$ region to the proximity
of the black dashed line produces a finite jump at the replica of
the main spin dispersion (marked by the green dotted line
in Fig.~\ref{fig:spectral_function}A)
the dispersion
of this replica, however, is indistinguishable from the principal
spin mode from the $l=0$ level in the particle sector. Around the
$k_{\rm F}$ point, the hierarchy breaks down so that all many-body
excitations have comparable amplitudes, and the spin mode regains
a singular peak obtained from the linear theory in \citep{Schoenhammer92,Voit93},
see the blue cut of $A\left(k,E\right)$ in Fig.~\ref{fig:spectral_function}C. 
\begin{figure}[h]
\centering
\includegraphics[width=1\columnwidth]{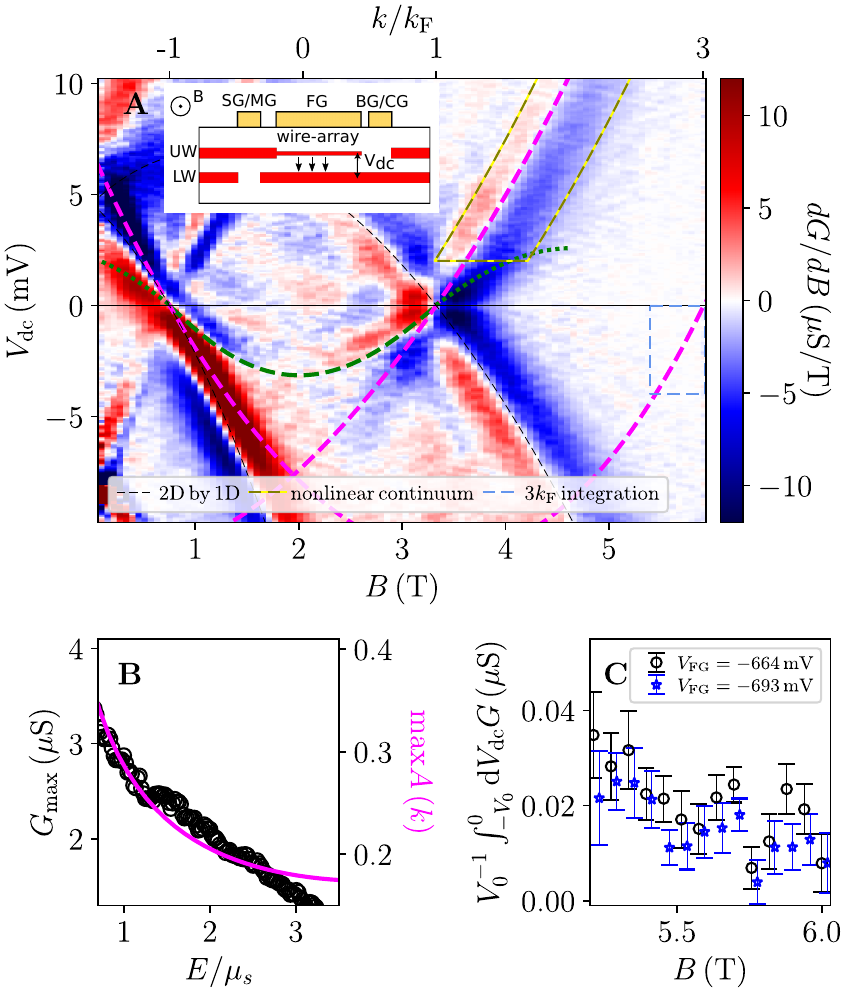}\caption{\textbf{A} The conductance $G(B,V_{\rm dc})$ measured for $V_{\rm FG}=-664$\,mV and presented as the $dG/dB$ derivative. The green and magenta dashed lines are the dispersions of spin and charge modes obtained from the full Lieb-Wu equations for $\gamma=1.25$, and corrected for capacitance using $c_{UL}=5.6\,{\rm mFm^{-2}}$ and $c_{UW}=4.7\,{\rm mFm^{-2}}$, see details in \cite{Vianez21}. The upper horizontal axis is the linear transformation of $B$ using the two crossing points with $V_{\rm dc}=0$ line as $B_{\rm lo}=0.75\,{\rm T}$ is $-k_{\rm F}$ and $B_{\rm hi}=3.33\,{\rm T}$ is $k_{\rm F}$. The inset is a schematic of the cross-section of our device. \textbf{B} Open black circles are the conductance along the charge mode in the region marked by the olive-yellow dashed line in \textbf{A}, for which $E/\mu_s$ is obtained as $V_{\rm dc}$ divided by the voltage of the minimum of the dashed-green parabola in \textbf{A}. The magenta line is the maximum of $A(k,E)$ in Fig.\ \ref{fig:spectral_function}A along the charge mode in the particle sector. \textbf{C} Integration over $V_{\rm dc}$ of the conductance $G$ within the light-blue dashed rectangle of height $V_0=4$\,mV in \textbf{A} as a function of $B$ for $V_{\rm FG}=-664$\,mV (black circles) and $V_{\rm FG}=-693$\,mV (blue stars). }\label{fig:transport_experiment}
\end{figure}

Here we turn to an experiment on tunneling spectroscopy in a 1D geometry (quantum wire) to search for the nonlinear features predicted by the Hubbard model in a real system with a screened Coulomb interaction. The design of our device \cite{Jompol09,Moreno16,Jin19} is based on a ${\rm GaAs/Al_{0.33}Ga_{0.67}As}$ double-quantum-well heterostructure. To amplify the 1D signal, the electrons in the upper well are confined to an array of 300 highly homogeneous wires of length $L=18\,{\rm \mu m}$ by applying a negative voltage $V_{\rm FG}$ on the finger gates, see the inset in Fig.\ \ref{fig:transport_experiment}A. The relative position of the Fermi energies of the wires and of the 2DEG in the lower well is shifted by applying a bias $V_{\rm dc}$ between the wells and the electron $k$-vector along the wire is shifted by $\Delta k=eBd/\hbar$ in the tunneling process by the Lorentz force from the in-plane magnetic field $B$ applied perpendicular to the wires, where $e$ is the electronic charge and $d=32$\,nm is the center-to-center distance between the two wells, providing together both energy and momentum resolution. The inter-well current $I$ measured for different $B$ and $V_{\rm dc}$ probes the spectral function of 1D electrons via its convolution with the 2D spectral function, see more details in \cite{SM}. An area of the upper well not covered by the finger gates always contributes parasitically to $I$, which we remove by repeating the measurement at $V_{\rm FG}=-750\,{\rm mV}$, when the wires are completely pinched off, and subtracting this signal from the 1D data at less negative $V_{\rm FG}$.

The conductance $G={\rm d}I/{\rm d}V_{\rm dc}$ for $V_{\rm FG}=-664\,{\rm mV}$ in the single 1D-subband regime with large density $n_{\rm 1D}\approx 45\,{\rm \mu m^{-1}}$ is presented in Fig.\ \ref{fig:transport_experiment}A. The data is visualized as the ${\rm d}G/{\rm d}B$ derivative to show the positions of the peaks as white lines between red and blue regions. The peak marked by the black dashed line is the 2D dispersion of the electrons in the lower well measured by the wires in the upper well. The pair of peaks marked by the green and magenta dashed lines have the same pattern as the maxima of the spectral function calculated in Fig.\ \ref{fig:spectral_function}A.
Fitting their dispersions by using the solution of the full Lieb-Wu equations \cite{LiebWu68} and $m_0=0.0525\,m_e$ \cite{Vianez23}, where $m_e$ is the free electron mass, we obtain a moderate value of the interaction parameter $\gamma=1.25$. Here the charge peak in $A(k,E)$ in the hole sector manifests as a minimum in ${\rm d}G/{\rm d}B$ due to peculiarities of the transport theory, which were already understood in the linear regime in \cite{Altland99}.

Furthermore, we find a signal consistent with the broad continuum of the nonlinear excitations predicted around the charge mode in the particle sector. In Fig.\ \ref{fig:transport_experiment}A it can be seen as a large asymmetry of this line (on the momentum scale of $\simeq k_{\rm F}$), see the area enclosed by the olive-yellow dashed line. The observed amplitude of this mode also decreases significantly in accordance with the Hubbard-model prediction, see the comparison in Fig.\ \ref{fig:transport_experiment}B and more analysis on the asymmetry in \cite{SM}. 

We also look for the many-body excitations around the $3k_{\rm F}$ point in the signal. The predicted amplitude of $A(k,E)$ around this point is $\sim 100$ times smaller than that around $k_{\rm F}$, bringing the expected amplitude of $G$ around $3k_{\rm F}$ just below the observed noise $\left|G\right|\sim 0.05\,{\rm \mu S}$. However, motivated by the relation  $n_k=\int_{-\infty}^0 {\rm d}E A(k,E)$, we integrate $G$ over $V_{\rm dc}$ in the light-blue dashed rectangle in Fig.\ \ref{fig:transport_experiment}A at each field and find a finite signal, the black circles in Fig.\ \ref{fig:transport_experiment}C. Repeating this measurement at a more negative $V_{\rm FG}=-693\,{\rm meV}$ (and a smaller $n_{\rm 1D}\approx 42\,{\rm \mu m^{-1}}$), we find a very similar pattern, the blue stars in Fig.\ \ref{fig:transport_experiment}C.
The non-monotonic dependence of the integrated $G$ cannot be explained by a still possible contribution from the far tail of the 2D signal in this region but, on the other hand, does not match the shape of $n_k$ around $3k_{\rm F}$ in Fig.\ \ref{fig:nk}A. The mismatch could be due to contribution of the next ($l=1,2$) levels of the hierarchy \cite{Tsyplyatyev15}, which were previously observed in spectroscopy \cite{Moreno16,Vianez21}, or due to a peculiarity of transport theory, which is still lacking for nonlinear Luttinger liquids. The latter may also be a reason for the absence of a quantitative match in the comparison in Fig.\ \ref{fig:transport_experiment}B.

In conclusion, we have developed a microscopic theory for the correlation function of the Hubbard model for $r_{\rm s}>1$ and have used it to show the stability of the nonlinear spin-charge separated modes and to investigate systematically their features. We have confirmed some of these predictions experimentally in a semiconductor quantum wires, indicating the reliability of the Hubbard-model prediction for a finite-range interaction. 

\emph{Acknowledgments}---For financial support, O.T.\ thanks the DFG (project No. 461313466) and C.J.B.F.\ thanks the UK EPSRC (grant no. EP/J01690X/1). We thank Ian Farrer and David Ritchie for growth of the heterostructure material and Jon Griffiths for the electron-beam lithography.

\emph{Data availability}---Experimental data presented in this work are available at the University of Cambridge data repository (http://...).

\emph{Author Contributions}---O.T. planned the project and performed all the calculations. Y.J., M.M.\ and W.K.T.\ fabricated the device and with C.J.B.F.\ performed the transport measurements shown. 
O.T.\ and C.J.B.F.\ analyzed the experimental data. O.T.\ wrote the manuscript\ with help from C.J.B.F..

\bibliographystyle{apsrev4-2}
\bibliography{citations}

\end{document}


\title{Supplementary material for \\A microscopic approach to nonlinear theory of spin-charge separation}

\maketitle
\tableofcontents{}

\section{Lieb-Wu solution}

The $N$-particle eigenstates of the 1D Hubbard model in Eq.\ (1)
of the main text were constructed by Lieb and Wu in \citep{LiebWu68}.
These eigenstate have the form of a superposition of plain waves according
to Bethe's hypothesis \citep{Bethe31}, which in second quantisation
$\left|\Psi\right\rangle =\sum_{\mathbf{j},\boldsymbol{\alpha}}\psi_{\mathbf{j}\boldsymbol{\alpha}}\prod_{i=1}^{N}c_{j_{i}\alpha_{i}}^{\dagger}\left|0\right\rangle $,
reads as

\begin{equation}
\psi_{\mathbf{j}\boldsymbol{\alpha}}=\sum_{Q}\left(-1\right)^{QO}A_{QO\boldsymbol{\alpha}}e^{iQ\mathbf{k}\cdot O\mathbf{j}},\label{eq:psi_jalpha}
\end{equation}
where $j_{1}\dots j_{N}=\mathbf{j}$ and $\alpha_{1}\dots\alpha_{N}=\boldsymbol{\alpha}$
are the coordinates and spin configurations of $N$ Fermi particles
with spin-1/2 on a chain of length $L$, $O$ is the permutation
that orders all $N$ coordinates so that 
\begin{equation}
Oj_{1}<\dots<Oj_{N},
\end{equation}
the charge quasimomenta associated with the orbital degrees of freedom
of every particle are $\mathbf{k}=k_{1}\dots k_{N}$, and $\sum_{Q}$
is the sum over all permutations $Q$ of $N$ quasimomenta $k_{j}$.
The amplitudes $A_{QO\boldsymbol{\alpha}}$ in this superposition
depend additionally on the spin configuration $O\boldsymbol{\alpha}$.
A way of constructing them using the Bethe hypothesis was proposed
in \citep{Yang67,Gaudin67}, producing another ``nested'' Bethe-ansatz
wave function for the spin degrees of freedom as 
\begin{equation}
A_{QO\boldsymbol{\alpha}}=\sum_{R}\prod_{1\leq l<m\leq M}\sqrt{-\frac{e^{i\left(Rq_{m}+Rq_{l}\right)}+1-2e^{iRq_{l}}}{e^{i\left(Rq_{m}+Rq_{l}\right)}+1-2e^{iRq_{m}}}}\prod_{l=1}^{M}\frac{i\frac{U}{2t}e^{iRq_{l}}}{\left(e^{iRq_{l}}-1\right)Qk_{x_{l}}+i\frac{U}{2t}}\prod_{j=1}^{x_{l}-1}\frac{\left(e^{iRq_{l}}-1\right)Qk_{j}+i\frac{U}{2t}e^{iRq_{l}}}{\left(e^{iRq_{l}}-1\right)Qk_{j}+i\frac{U}{2t}},\label{eq:AQOalpha}
\end{equation}
where $x_{1}\dots x_{M}=\mathbf{x}$ are the coordinates of $M$ spins
$\uparrow$ in the configuration $O\boldsymbol{\alpha}$ of all spins
of $N$ particles, $q_{1}\dots q_{M}=\mathbf{q}$ are the spin quasimomenta
associated with these $M$ spins $\uparrow$, and $\sum_{R}$ is the
sum over all permutations $R$ of $M$ quasimomenta $q_{j}$. Note
that this spin wave function is written in the coordinate representation,
in which the normalisation factors are defined in this work. The expression
in Eq.~(\ref{eq:AQOalpha}), used as the starting point in this work,
is obtained by substitution of the algebraic-to-coordinate mapping
for the spin quasimomenta as 
\begin{equation}
\lambda_{l}=-\frac{iU}{4t}\frac{e^{iq_{l}}+1}{e^{iq_{l}}-1}
\end{equation}
in the original expression in the algebraic representation in \citep{Yang67,Gaudin67}.
The low-particle-density limit, $N\ll L$, is already taken in Eq.~(\ref{eq:AQOalpha}).
For a large lattice filling, the charge quasimomenta have to be changed
as $Qk_{j}\rightarrow\sin Qk_{j}$ in this expression.

Application of the periodic boundary condition to the wave function
in Eq.~(\ref{eq:psi_jalpha}) solves the eigenvalue problem by a
set of coupled nonlinear equations, the Lieb-Wu equations \citep{LiebWu68},
\begin{equation}
Lk_{j}=\frac{1}{i}\sum_{l=1}^{M}\log\left(1+\frac{1}{\frac{1}{e^{iq_{l}}-1}-\frac{2it}{U}k_{j}}\right)+2\pi I_{j},\label{eq:LiebWu_Ij}
\end{equation}
\begin{equation}
\frac{1}{i}\sum_{j=1}^{N}\log\left(1+\frac{1}{\frac{1}{e^{iq_{m}}-1}-\frac{2it}{U}k_{j}}\right)=\frac{1}{i}\sum_{l=1\neq m}^{M}\log\left(-\frac{e^{i\left(q_{l}+q_{m}\right)}+1-2e^{iq_{m}}}{e^{i\left(q_{l}+q_{m}\right)}+1-2e^{iq_{l}}}\right)+2\pi J_{m},\label{eq:LiebWu_Jm}
\end{equation}
where $N$ non-equal integers $I_{j}$ and $M$ non-equal integers
$J_{m}$ define the solution for the corresponding charge and spin
quasimomenta $k_{j}$ and $q_{j}$ for a given value of the interaction
strength $U/t$. This solution gives the eigenenergy of the corresponding
$N$-particle state as $E=t\sum_{j}k_{j}^{2}/2$ and its total momentum
as $P=\sum_{j}k_{j}$. The low-particle-density limit, $N\ll L$,
is already taken in these Lieb-Wu equations. The expressions for arbitrary
density are obtained via the $k_{j}\rightarrow\sin k_{j}$ substitution
in the r.h.s. of Eq.~(\ref{eq:LiebWu_Ij}) and in the whole of Eq.~(\ref{eq:LiebWu_Jm}).

\subsection{$t/U$ expansion of eigenvalue equations}

The Lieb-Wu equations (\ref{eq:LiebWu_Ij},\ref{eq:LiebWu_Jm}) can
be solved explicitly for large $U$ by expanding them in the Taylor
series in $t/U$. The leading term in such an expansion is obtained
by taking the $U\rightarrow\infty$ limit of Eqs.~(\ref{eq:LiebWu_Ij}, \ref{eq:LiebWu_Jm}),
which gives \citep{Ogata90}
\begin{equation}
Lk_{j}^{0}=P_{\rm s}+2\pi I_{j},\label{eq:LiebWu_Ij0}
\end{equation}
\begin{equation}
Nq_{m}^{0}=2\sum_{l\neq m}\varphi_{ml}+2\pi J_{m},\label{eq:LiebWu_Jm0}
\end{equation}
where 
\begin{equation}
e^{i2\varphi_{ml}}=-\frac{e^{i\left(q_{l}^{0}+q_{m}^{0}\right)}+1-2e^{iq_{m}^{0}}}{e^{i\left(q_{l}^{0}+q_{m}^{0}\right)}+1-2e^{iq_{l}^{0}}},\label{eq:phi_lm}
\end{equation}
are the two-spinon scattering phases, the superscript $0$ was added
to both charge $k_{j}^{0}$ and spin $q_{j}^{0}$ quasimomenta to
mark the zeroth-order term in the $t/U$ Taylor series for them. The
total spin momentum
\begin{equation}
P_{\rm s}=\sum_{l}q_{l}^{0}
\end{equation}
becomes a quantum number in this limit since the spin part in Eq.~(\ref{eq:LiebWu_Jm0})
decouples completely, becoming an independent set of $M$ nonlinear
equation for $q_{j}^{0}$ only, which are exactly the same as the
Bethe equations for the antiferromagnetic Heisenberg chain \citep{Bethe31}.
Once the solution for $q_{m}^{0}$ is found from Eq.~(\ref{eq:LiebWu_Jm0})
for a set of $J_{m}$, each of the remaining $N$ equations for $k_{j}$
in Eq.~(\ref{eq:LiebWu_Jm0}) becomes just an independent single-particle
quantisation condition, which is solved immediately as 
\begin{equation}
k_{j}^{0}=\frac{P_{\rm s}+2\pi J_{m}}{L}.\label{eq:k_j0}
\end{equation}

To find the next term in the $t/U$ expansion, the Lieb-Wu equations
(\ref{eq:LiebWu_Ij}, \ref{eq:LiebWu_Jm}) have to be expanded in a
Taylor series in $t/U$ around the $U=\infty$ point. In order to
evaluate the linear term for the spin and charge quasimomenta, they
are expanded up to linear order in $t/U$ around the solutions
of Eqs.~(\ref{eq:LiebWu_Ij0}, \ref{eq:LiebWu_Jm0}),

\begin{align}
k_{j} & =k_{j}^{0}+\frac{t}{U}k_{j}^{1},\label{eq:kj_01}\\
q_{m} & =q_{m}^{0}+\frac{t}{U}q_{m}^{1}.\label{eq:qm_01}
\end{align}
Then, these expansions are substituted into Eqs.~(\ref{eq:LiebWu_Ij}, \ref{eq:LiebWu_Jm})
and they are, in turn, expanded up to linear order in $t/U$, in
which the equations for the quasimomenta become linear in $k_{j}^{1}$
and $q_{j}^{1}$, 
\begin{equation}
Lk_{j}^{1}=\sum_{l}q_{l}^{1}+4k_{j}^{0}\sum_{l}\left(\cos q_{l}^{0}-1\right),\label{eq:LiebWu_Ij1}
\end{equation}
\begin{equation}
Nq_{m}^{1}=4\sum_{l\neq m}\frac{q_{m}^{1}\left(1-\cos q_{l}^{0}\right)-q_{l}^{1}\left(1-\cos q_{m}^{0}\right)}{\left(e^{iq_{m}^{0}}+e^{-iq_{l}^{0}}-2\right)\left(e^{iq_{l}^{0}}+e^{-iq_{m}^{0}}-2\right)}-4\left(\cos q_{m}^{0}-1\right)\sum_{j}k_{j}^{0}.\label{eq:LiebWu_Jm1}
\end{equation}

Summation of Eq.~(\ref{eq:LiebWu_Jm1}) over $m$ makes the first
term on its r.h.s.\ vanish, giving
\begin{equation}
\sum_{m}q_{m}^{1}=-\frac{4}{N}\sum_{l}\left(\cos q_{l}^{0}-1\right)\sum_{j}k_{j}^{0},
\end{equation}
since the summand is asymmetric w.r.t.\ exchange of the summation indices
$l$ and $m$. Substitution of the last expression in Eq.~(\ref{eq:LiebWu_Ij1})
gives the first order for the charge quasimomenta $k_{j}^{1}$ explicitly
in terms of the solutions of Eqs.~(\ref{eq:LiebWu_Ij0}, \ref{eq:LiebWu_Jm0})
only,
\begin{equation}
k_{j}^{1}=\frac{4}{L}\left(k_{j}^{0}-\frac{P^{0}}{N}\right)\sum_{l}\left(\cos q_{l}^{0}-1\right),\label{eq:k1j_full}
\end{equation}
where we note that the quantum number $P^{0}=\sum_{j}k_{j}^{0}\equiv P$
is independent of $U$ due to the translational invariance.

The first order for the spin quasimomenta $q_{m}^{1}$ can be found
from Eq.~(\ref{eq:LiebWu_Jm1}) via matrix inversion as
\begin{equation}
q_{m}^{1}=\left[\hat{Q}_{0}^{-1}\mathbf{v}\right]_{m},\label{eq:q1m}
\end{equation}
where $\mathbf{v}$ is the second term on the r.h.s.\ of Eq.~(\ref{eq:LiebWu_Jm1})
written in the vectorial form
\begin{equation}
v_{m}=\left(1-\cos q_{m}^{0}\right)P,
\end{equation}
and the matrix 
\begin{equation}
Q_{ml}^{0}=\begin{cases}
N-\sum_{k=1\neq m}^{M}\frac{4\left(1-\cos q_{k}^{0}\right)}{\left(e^{iq_{m}^{0}}+e^{-iq_{k}^{0}}-2\right)\left(e^{iq_{k}^{0}}+e^{-iq_{m}^{0}}-2\right)}, & m=l,\\
\frac{4\left(1-\cos q_{m}^{0}\right)}{\left(e^{iq_{m}^{0}}+e^{-iq_{l}^{0}}-2\right)\left(e^{iq_{l}^{0}}+e^{-iq_{m}^{0}}-2\right)}, & m\neq l,
\end{cases}\label{eq:Q_ml0}
\end{equation}
is the same matrix whose determinant gives the normalisation of the
Bethe wave functions, $Z^{2}={\rm det}\hat{Q}_{0}$ \citep{Gaudin81}.
Since the normalisation factor is always finite for the physically meaningful
states, $\det\hat{Q}_{0}$ is also finite, making the matrix $\hat{Q}_{0}$
invertible for all eigenstates of the Heisenberg chain.

In the thermodynamic limit, $N,L\gg1$, the contribution of $P^{0}/N$
in Eq.~(\ref{eq:k1j_full}) becomes subleading for physically relevant
momenta within a few-Fermi-momenta range so it can be neglected, 
\begin{equation}
k_{j}^{1}=\frac{4}{L}k_{j}^{0}\sum_{l}\left(\cos q_{l}^{0}-1\right).\label{eq:k_j1}
\end{equation}
The result in the last equation is presented in Eq.~(2) of the main
paper. The result in Eq.~(\ref{eq:q1m}) is presented in Eq.~(3)
of the main paper.

\subsection{$t/U$ expansion of wave function}

The Lieb-Wu wave function (\ref{eq:psi_jalpha}) simplifies for large
$U$, which manifests in its expansion in the Taylor series in $t/U$.
In the $U=\infty$ limit, the amplitude in Eq.~(\ref{eq:AQOalpha})
becomes independent of the charge quasimomenta $k_{j}^{0}$ and their
permutation $Q$, 
\begin{equation}
A_{QO\boldsymbol{\alpha}}=\sum_{R}e^{i\sum_{l<m}\varphi_{R_{l},R_{m}}+i\sum_{l}q_{l}^{0}x_{l}},\label{eq:AQOalpha0}
\end{equation}
where the two-spinon scattering phase $\varphi_{lm}$ is given by
Eq.~(\ref{eq:phi_lm}). Then, the ordering permutation $O$ of spin
configurations $\boldsymbol{\alpha}$ is absorbed in the relabelling
of $\boldsymbol{\alpha}$ under the sum over all $\boldsymbol{\alpha}$,
this amplitude can be taken out of the sum over the permutation $Q$
in Eq.~(\ref{eq:psi_jalpha}), the ordering permutation $O$ of the
charge coordinates $\mathbf{j}$ is absorbed in the relabelling of
the permutation $Q$ under the sum over all $Q$, and the whole Lieb-Wu
wave function factorises in this limit as \citep{Ogata90} 
\begin{equation}
\psi_{\mathbf{j}\boldsymbol{\alpha}}^{0}=\psi_{\mathbf{j}}^{0c}\cdot\psi_{\boldsymbol{\alpha}}^{0s},\label{eq:psi_jalpha0}
\end{equation}
where 
\begin{equation}
\psi_{\mathbf{j}}^{0c}=L^{-N}\sum_{Q}\left(-1\right)^{Q}e^{iQ\mathbf{k}^{0}\cdot \mathbf{j}}\label{eq:psi_j0c}
\end{equation}
is a Slater determinant that depends only on the charge coordinates
$\mathbf{j}$, 
\begin{equation}
\psi_{\boldsymbol{\alpha}}^{0s}=Z^{-1}\sum_{R}e^{i\sum_{l<m}\varphi_{R_{l},R_{m}}+i\sum_{l}Rq_{l}^{0}x_{l}}\label{eq:psi_alpha0s}
\end{equation}
is the Bethe wave function for the antiferromagnetic spin chain \citep{Bethe31}
that depends only on the spin configurations $\boldsymbol{\alpha}$,
and $Z=\sqrt{\det\hat{Q}_{0}}$ is the Gaudin normalisation factor
\citep{Gaudin81}. Here each of the factorised charge and spin
wave functions are already normalised to unity so that the whole Lieb-Wu
wave function~(\ref{eq:psi_jalpha0}) is also normalised to unity.

In the representation of second quantisation, such a wave function factorised
into the spin and charge sectors in Eq.~(\ref{eq:psi_jalpha0})
can be written as a direct product,
\begin{equation}
|\Psi^{0}\rangle=|\Psi_{\rm c}^{0}\rangle\otimes|\Psi_{\rm s}^{0}\rangle,\label{eq:Psi0}
\end{equation}
in which the charge part is 
\begin{equation}
\left|\Psi_{\rm c}^{0}\right\rangle =\frac{1}{L^{\frac{N}{2}}N!^{\frac{1}{2}}}\sum_{Q,\mathbf{j}}\left(-1\right)^{Q}e^{iQ\mathbf{k}^{0}\cdot O\mathbf{j}}a_{j_{1}}^{\dagger}\cdots a_{j_{N}}^{\dagger}\left|0\right\rangle \label{eq:Psi_c0}
\end{equation}
and the spin part is 
\begin{equation}
\left|\Psi_{\rm s}^{0}\right\rangle =\frac{1}{Z}\sum_{R,x_{1}<\dots<x_{M}}e^{i\sum_{l<m}\varphi_{R_{l}R_{m}}+iR\mathbf{q}\cdot\mathbf{x}}S_{x_{1}}^{+}\cdots S_{x_{M}}^{+}\left|\Downarrow\right\rangle ,\label{eq:Psi_s0}
\end{equation}
where $a_{j}^{\pm}$ are the spinless ladder operators obeying Fermi
statistics on the same 1D lattice of length $L$ and $S_{j}^{\pm}$
are the spin-1/2 flip operators on the spin chain of length $N$ formed
by the spin degree of freedom of $N$ particles. These operators can
be recombined into the original electron operators $c_{j\alpha}^{\pm}$
of the Hubbard model in Eq.~(1) of the main text by introducing an
insertion (deletion) operator of a site in the spin-down state at
a given position $x$ on the spin chain $I_{x}$ ($D_{x}$) as 
\begin{align}
c_{j\uparrow}^{\dagger} & =a_{j}^{\dagger}S_{x}^{+}I_{x},\\
c_{j\downarrow}^{\dagger} & =a_{j}^{\dagger}I_{x},\\
c_{j\uparrow} & =a_{j}D_{x}S_{x}^{-},\label{eq:c_jup}\\
c_{j\downarrow} & =a_{j}D_{x}S_{x}^{-}S_{x}^{+}.
\end{align}
In the r.h.s.\ of these equations, the $a_{j}^{\pm}$ and $S_{x}^{\pm}$
operators obey the Fermi and spin commutation rules but the $I_{x}$
and $D_{x}$ operators do not. In general, these insertion and deletion
operators also do not commute with the spin operators.

Like the Lieb-Wu equations (\ref{eq:LiebWu_Ij}, \ref{eq:LiebWu_Jm})
in the previous subsection, the Lieb-Wu wave function (\ref{eq:psi_jalpha})
can also be expanded in a Taylor series in $t/U$ around the $U=\infty$
point. Substitution of the linear expansion for the charge and spin
quasimomenta around the $U=\infty$ point from Eqs.~(\ref{eq:kj_01}, \ref{eq:qm_01})
into Eq.~(\ref{eq:psi_jalpha}) and expansion of the resulting expression
up to linear order in $t/U$ gives 
\begin{equation}
\psi_{\mathbf{j}\boldsymbol{\alpha}}=\psi_{\mathbf{j}\boldsymbol{\alpha}}^{0}+t/U\psi_{\mathbf{j}\boldsymbol{\alpha}}^{1},
\end{equation}
where the zeroth term $\psi_{\mathbf{j}\boldsymbol{\alpha}}^{0}$
is factorised into the spin and charge sectors as given by Eq.~(\ref{eq:psi_jalpha0})
and the linear term has three contributions, 
\begin{equation}
\psi_{\mathbf{j}\boldsymbol{\alpha}}^{1}=\psi_{\mathbf{j}\boldsymbol{\alpha}}^{1c}+\psi_{\mathbf{j}\boldsymbol{\alpha}}^{1sc}+\psi_{\mathbf{j}\boldsymbol{\alpha}}^{1s}.\label{eq:psi_jalpha1}
\end{equation}
The first and the third terms in $\psi_{\mathbf{j}\boldsymbol{\alpha}}^{1}$
come from the linear terms in charge and spin quasimomenta in Eqs.~(\ref{eq:kj_01}, \ref{eq:qm_01}), respectively, when the amplitude in Eq.~(\ref{eq:AQOalpha}) has the
form of the $U=\infty$ limit in Eq.~(\ref{eq:AQOalpha0}). These
terms do not contain any mixing between the charge and spin variables
so that both $\psi_{\mathbf{j}\boldsymbol{\alpha}}^{1s}$ and $\psi_{\mathbf{j}\boldsymbol{\alpha}}^{1c}$
are still factorised in the spin and charge sectors
\begin{equation}
\psi_{\mathbf{j}\boldsymbol{\alpha}}^{1s}=\psi_{\mathbf{j}}^{0c}\cdot\frac{1}{Z}\sum_{R}e^{i\sum_{l<m}\varphi_{R_{l}R_{m}}+iR\mathbf{q}^{0}\cdot\mathbf{x}}\left[2\sum_{l<m}\frac{Rq_{l}^{1}\left(1-\cos Rq_{m}^{0}\right)-Rq_{m}^{1}\left(1-\cos Rq_{m}^{0}\right)}{\left(e^{iRq_{l}^{0}}+e^{-iRq_{m}^{0}}-2\right)\left(e^{-iRq_{l}^{0}}+e^{iRq_{m}^{0}}-2\right)}+iR\mathbf{q}^{1}\cdot\mathbf{x}\right],
\end{equation}
\begin{equation}
\psi_{\mathbf{j}\boldsymbol{\alpha}}^{1c}=\left[\frac{i}{L^{\frac{N}{2}}N!^{\frac{1}{2}}}\sum_{Q}\left(-1\right)^{Q}e^{iQ\mathbf{k}^{0}\cdot O\mathbf{j}}Q\mathbf{k}^{1}\cdot O\mathbf{j}\right]\cdot\psi_{\boldsymbol{\alpha}}^{0s},
\end{equation}
where $\psi_{\mathbf{j}}^{0c}$ and $\psi_{\boldsymbol{\alpha}}^{0s}$
are given in Eqs.~(\ref{eq:psi_jalpha0}) and (\ref{eq:psi_j0c})
respectively. The second term in Eq.~(\ref{eq:psi_jalpha1}) comes from the
linear expansion of Eq.~(\ref{eq:AQOalpha}) representing the linearised
version of the original mixing between the spin and charge degrees
of freedom in the Lieb-Wu wave function,
\begin{equation}
\psi_{\mathbf{j}\boldsymbol{\alpha}}^{1sc}=-\frac{4}{L^{\frac{N}{2}}N!^{\frac{1}{2}}Z}\sum_{Q,R}\left(-1\right)^{Q}e^{iQ\mathbf{k}^{0}\cdot O\mathbf{j+}i\sum_{l<m}\varphi_{R_{l}R_{m}}+iR\mathbf{q}\cdot\mathbf{x}}\sum_{m'}\left[\frac{1-e^{iRq_{m'}^{0}}}{2}Qk_{x_{m'}}^{0}+\left(1-\cos Rq_{m'}^{0}\right)\sum_{j'=1}^{x_{m'}}Qk_{j'}^{0}\right].
\end{equation}

In the representation of second quantisation, the expansion of
the Lieb-Wu wave function up to linear order in $t/U$ reads as
\begin{equation}
\left|\Psi\right\rangle =\left|\Psi^{0}\right\rangle +t/U\left|\Psi^{1}\right\rangle .\label{eq:Psi01}
\end{equation}
The zeroth term is already given by Eq.~(\ref{eq:Psi0}). The three
contributions to the linear term of the wave function in Eqs.~(\ref{eq:psi_jalpha1})
can be written in this representation as 
\begin{equation}
\left|\Psi^{1}\right\rangle =\left|\Psi_{\rm c}^{1}\right\rangle +\left|\Psi_{sc}^{1}\right\rangle +\left|\Psi_{\rm s}^{1}\right\rangle ,\label{eq:Psi1}
\end{equation}
where 
\begin{equation}
\left|\Psi_{\rm c}^{1}\right\rangle =\frac{i}{L^{\frac{N}{2}}N!^{\frac{1}{2}}}\sum_{Q,\mathbf{j}}\left(-1\right)^{Q}e^{iQ\mathbf{k}^{0}\cdot O\mathbf{j}}Q\mathbf{k}^{1}\cdot O\mathbf{j}a_{j_{1}}^{\dagger}\cdots a_{j_{N}}^{\dagger}\left|0\right\rangle \otimes\left|\Psi_{\rm s}^{0}\right\rangle ,\label{eq:Psi1c}
\end{equation}
\begin{multline}
\left|\Psi_{sc}^{1}\right\rangle =-\frac{4i}{L^{\frac{N}{2}}N!^{\frac{1}{2}}Z}\sum_{Q,R,\mathbf{j},\mathbf{x}}\left(-1\right)^{Q}e^{iQ\mathbf{k}^{0}\cdot O\mathbf{j}+i\sum_{l<m}\varphi_{R_{l}R_{m}}+iR\mathbf{q}\cdot\mathbf{x}}\\
\sum_{m'}\left[\frac{1-e^{iRq_{m'}^{0}}}{2}Qk_{x_{m'}}^{0}+\left(1-\cos Rq_{m'}^{0}\right)\sum_{j'=1}^{x_{m_{'}}}Qk_{j'}^{0}\right]a_{j_{1}}^{\dagger}\cdots a_{j_{N}}^{\dagger}\left|0\right\rangle \otimes S_{x_{1}}^{+}\cdots S_{x_{M}}^{+}\left|\Downarrow\right\rangle ,\label{eq:Psi1sc}
\end{multline}
\begin{multline}
\left|\Psi_{\rm s}^{1}\right\rangle =\left|\Psi_{\rm c}^{0}\right\rangle \otimes\frac{1}{Z}\sum_{R,\mathbf{x}}e^{i\sum_{l<m}\varphi_{R_{l}R_{m}}+iR\mathbf{q}^{0}\cdot\mathbf{x}}\\
\left[i2\sum_{l<m}\frac{Rq_{l}^{1}\left(1-\cos Rq_{m}^{0}\right)-Rq_{m}^{1}\left(1-\cos Rq_{m}^{0}\right)}{\left(e^{iRq_{l}^{0}}+e^{-iRq_{m}^{0}}-2\right)\left(e^{-iRq_{l}^{0}}+e^{iRq_{m}^{0}}-2\right)}+iR\mathbf{q}^{1}\cdot\mathbf{x}\right]S_{x_{1}}^{+}\cdots S_{x_{M}}^{+}\left|\Downarrow\right\rangle .\label{eq:Psi1s}
\end{multline}
Here $\left|\Psi_{\rm c}^{0}\right\rangle $ and $\left|\Psi_{\rm s}^{0}\right\rangle $
are the factorised charge and spin parts of the Lieb-Wu wave function
in the $U=\infty$ limit given by Eqs.~(\ref{eq:Psi_c0}, \ref{eq:Psi_s0})
and the sum over the spin coordinates $\mathbf{x}$ always runs only
over the ordered set of coordinates, $x_{1}<\dots<x_{M}$ , which
is not specified explicitly in what follows for brevity. 

\section{Algebraic Bethe ansatz}

The spin part of the wave function in the $U=\infty$ limit in Eq.~(\ref{eq:psi_alpha0s})
is the same as the Bethe wave function for the antiferromagnetic Heisenberg
model in 1D \citep{Bethe31}. The Bethe wave function in the coordinate
representation in Eq.~(\ref{eq:psi_alpha0s}) is not factorised in
terms of the single-spin states, making calculation of the matrix elements
impossible. However, an algebraic representation of the Bethe wave
function was invented in \citep{Sklyanin79} to factorise it in terms
of operators with specific commutation relations, which can be used
for the analytical calculations of matrix elements. Here, we briefly
introduce this algebraic representation, which will be used later
for dealing with the spin part of matrix elements of the Hubbard model.
It is more convenient to do this algebraic construction for a more
general model, the XXZ spin model, 
\begin{equation}
H=\sum_{j=1}^{N}\left(\frac{S_{j}^{+}S_{j+1}^{-}+S_{j}^{-}S_{j+1}^{+}}{2}+\Delta S_{j}^{z}S_{j+1}^{z}\right),\label{eq:H_XXZ}
\end{equation}
for which the eigenstates are the same as in Eq.~(\ref{eq:psi_alpha0s})
where the spin quasimomenta $q_{j}^{0}$ satisfy almost the same set
of equations~(\ref{eq:LiebWu_Jm0}). The only difference is in the
two-magnon scattering phase,
\begin{equation}
e^{i2\varphi_{ml}}=-\frac{e^{i\left(q_{l}^{0}+q_{m}^{0}\right)}+1-2\Delta e^{iq_{m}^{0}}}{e^{i\left(q_{l}^{0}+q_{m}^{0}\right)}+1-2\Delta e^{iq_{l}^{0}}}\label{eq:phi_lm_XXZ}
\end{equation}
that needs to be used for the model in Eq.~(\ref{eq:H_XXZ}) instead
of the scattering phase in Eq.~(\ref{eq:phi_lm}). For $\Delta=1$
the XXZ model in Eq.~(\ref{eq:H_XXZ}) becomes the antiferromagnetic
Heisenberg model $H=\sum_{j}\mathbf{S}_{i}\cdot\mathbf{S}_{j+1}$
and the two-spinon scattering phase in Eq.~(\ref{eq:phi_lm_XXZ})
becomes that of the antiferromagnetic Heisenberg model in Eq.~(\ref{eq:phi_lm}).

In this work we follow the notations of the book in \citep{Korepin_book}.
The $M$-spinon eigenfunction of the XXZ model can be represented
by the algebraic Bethe operators generated by the Yang-Baxter equation
\citep{Yang67,Baxter72} as 
\begin{equation}
\left|\mathbf{u}\right\rangle =\prod_{m=1}^{M}C\left(u_{j}\right)\left|\Downarrow\right\rangle ,\label{eq:Bethe_wavefunction_ABA}
\end{equation}
where $u_{j}$ are $M$ complex parameters corresponding to the $M$
spin quasimomenta $q_{j}$, $\left|\Downarrow\right\rangle $ is the
``vacuum state'' of the spin chain, and $C\left(u\right)$ is
a matrix element of the monodromy matrix 
\begin{equation}
T\left(u\right)=\left(\begin{array}{cc}
A\left(u\right) & B\left(u\right)\\
C\left(u\right) & D\left(u\right)
\end{array}\right).\label{eq:T_matrix}
\end{equation}
This matrix $T\left(u\right)$ is defined in an auxiliary $2\times2$
space, it is a function of a complex parameter $u$, and its four
entries are operators that act in the space of $N$ spins forming
the chain. When this matrix is a solution of the Yang-Baxter equation,
\begin{equation}
R\left(u-v\right)\left(T\left(u\right)\otimes T\left(v\right)\right)=\left(T\left(v\right)\otimes T\left(u\right)\right)R\left(u-v\right),\label{eq:YangBaxter_equation}
\end{equation}
the $M$-body scattering matrices factorise into products of only
two-body scattering matrices. This equation is defined by the $R$-matrix
that acts on a $4\times4$ tensor product space $V_{1}\otimes V_{2}$,
where $V_{1}$ and $V_{2}$ are two-element subspaces. For the XXZ
model in Eq.~(\ref{eq:H_XXZ}) the $R$-matrix is \citep{Korepin_book}
\begin{equation}
R\left(u\right)=\left(\begin{array}{cccc}
1\\
 & b\left(u\right) & c\left(u\right)\\
 & c\left(u\right) & b\left(u\right)\\
 &  &  & 1
\end{array}\right),\label{eq:R_matrix}
\end{equation}
where 
\begin{equation}
b\left(u\right)=\frac{\sinh\left(u\right)}{\sinh\left(u+2\eta\right)},\quad c\left(u\right)=\frac{\sinh\left(2\eta\right)}{\sinh\left(u+2\eta\right)},\label{eq:b(u), c(u)}
\end{equation}
and $\eta$ is a real parameter corresponding to $\Delta$ in the
XXZ model. Note that this $R$-matrix also satisfies the Yang-Baxter
equation defined by itself, 
\begin{equation}
R_{12}\left(u_{1}-u_{2}\right)R_{13}\left(u_{1}\right)R_{23}\left(u_{2}\right)=R_{23}\left(u_{2}\right)R_{13}\left(u_{1}\right)R_{12}\left(u_{1}-u_{2}\right).
\end{equation}

For a chain consisting of only one spin, the solution of Eq.~(\ref{eq:YangBaxter_equation})
can be constructed by identifying one two-element subspace of the
$R$-matrix in Eq.~(\ref{eq:R_matrix}) with the two-state spin-1/2
space of a single lattice spin on site $j$. Then, this $R$-matrix
becomes the quantum version of the Lax matrix \cite{Lax68} for such a single-site chain, $L_{j}=R_{1j}$, in which the other two-element subspace plays
the role of the auxiliary $2\times2$ space of the $T$-matrix in
Eq.~(\ref{eq:T_matrix}). In this auxiliary subspace the single-site
Lax matrix reads as 
\begin{equation}
L_{j}\left(u\right)=\left(\begin{array}{cc}
\frac{\cosh\left(u+\eta2S_{j}^{z}\right)}{\cosh\left(u-\eta\right)} & -i\frac{\sinh\left(2\eta\right)2S_{j}^{-}}{\cosh\left(u-\eta\right)}\\
-i\frac{\sinh\left(2\eta\right)2S_{j}^{+}}{\cosh\left(u-\eta\right)} & \frac{\cosh\left(u-\eta2S_{j}^{z}\right)}{\cosh\left(u-\eta\right)}
\end{array}\right).\label{eq:L_j}
\end{equation}
Further, the monodromy matrix for the whole chain consisting of $N$
spins is constructed as
\begin{equation}
T\left(u\right)=\prod_{j=1}^{N}L_{j}\left(u\right),\label{eq:T_XXZ}
\end{equation}
providing a definition of the algebraic Bethe ansatz operators in
terms of the physical spin operators of the model in Eq.~(\ref{eq:H_XXZ}).
By construction, the $T$-matrix in Eq.~(\ref{eq:T_XXZ}) satisfies
the Yang-Baxter equation (\ref{eq:YangBaxter_equation}) with the
$R$-matrix in Eq.~(\ref{eq:R_matrix}), see details in \citep{Korepin_book}.

The entries of the Yang-Baxter equation (\ref{eq:YangBaxter_equation})
in the $4\times4$ space with the $R$-matrix of the XXZ model in
Eq.~(\ref{eq:R_matrix}) give the explicit form of the commutations
relation between all four Bethe ansatz operators $A\left(u\right)$,
$B\left(u\right)$, $C\left(u\right)$, and $D\left(u\right)$ defined
in Eq.~(\ref{eq:T_matrix}). The commutation relations that we will
need later are 
\begin{equation}
\left[B_{u},C_{v}\right]=\frac{c\left(u-v\right)}{b\left(u-v\right)}\left(A_{u}D_{v}-A_{v}D_{u}\right),\label{eq:BuCv}
\end{equation}
\begin{equation}
A_{u}C_{v}=\frac{1}{b\left(u-v\right)}C_{v}A_{u}-\frac{c\left(u-v\right)}{b\left(u-v\right)}C_{u}A_{v},\label{eq:AuCv}
\end{equation}
\begin{equation}
D_{u}C_{v}=\frac{1}{b\left(u-v\right)}C_{v}D_{u}-\frac{c\left(v-u\right)}{b\left(v-u\right)}C_{u}D_{v},\label{eq:DuCv}
\end{equation}
\begin{equation}
\left[A_{u},D_{v}\right]=\frac{c\left(u-v\right)}{b\left(u-v\right)}\left(C_{v}B_{u}-C_{u}B_{v}\right),\label{eq:AuDv}
\end{equation}
where the subscripts $u$ and $v$ were introduced for brevity, \emph{e.g.},
$A_{u}\equiv A\left(u\right)$. 

The transition matrix $\tau\left(u\right)$ is given by the trace
of the $T$-matrix in the algebraic approach,
\begin{equation}
\tau\left(u\right)=A\left(u\right)+D\left(u\right).
\end{equation}
This operator also commutes with itself for different values of $u$,
\emph{i.e.}, $\left[\tau\left(u\right),\tau\left(v\right)\right]=0$
for $u\neq v$, being a linear superposition of all the conserved
quantities of the problem, including the XXZ model. Therefore, if
a state $\left|\mathbf{u}\right\rangle $, parametrised by a set of
$u_{j}$, is an eigenstates of $\tau\left(u\right)$ it is also an
eigenstates of the Hamiltonian in Eq.~(\ref{eq:H_XXZ}),
\begin{equation}
\tau\left(u\right)\left|\mathbf{u}\right\rangle =\mathcal{T}_{u}\left|\mathbf{u}\right\rangle ,
\end{equation}
where $\mathcal{T}_{u}$ is the corresponding eigenvalue of the transition
matrix. Solution of this eigenvalue problem imposes a constraint on
the $M$ parameters $u_{j}$ as \citep{Korepin_book}
\begin{equation}
\frac{a_{m}}{d_{m}}=\prod_{l=1\neq m}^{M}\frac{b_{ml}}{b_{lm}},\label{eq:Bethe_equations_ABA_v0}
\end{equation}
where $a_{m}$ and $d_{m}$ are the vacuum eigenvalues of the $A_{u}$
and $D_{u}$ operators, $A_{u}\left|\Downarrow\right\rangle =a_{u}\left|\Downarrow\right\rangle $
and $D_{u}\left|\Downarrow\right\rangle =d_{u}\left|\Downarrow\right\rangle $,
and the subscripts were introduced further as $a_{m}\equiv a\left(u_{m}\right)$,
$d_{m}\equiv d\left(u_{m}\right)$, $b_{ml}\equiv b\left(u_{m}-u_{l}\right)$,
and $b_{mu}\equiv b\left(u_{m}-u\right)$ for brevity. When this constraint
is obeyed, \emph{i.e.}, a particular set of $u_{j}$ gives an eigenstate
of $\tau\left(u\right)$, the corresponding eigenvalue of $\tau\left(u\right)$
is 
\begin{equation}
\mathcal{T}_{u}=a_{u}\prod_{m=1}^{M}\frac{1}{b_{um}}+d_{u}\prod_{m=1}^{M}\frac{1}{b_{mu}}.\label{eq:Tau_u_v0}
\end{equation}

The vacuum eigenvalues $a_{m}$ and $d_{m}$ are evaluated straightforwardly
using the expressions constructed in Eqs.~(\ref{eq:L_j},\nolinebreak\space\ref{eq:T_XXZ})
and the properties of Pauli matrices as 
\begin{equation}
a_{u}=\frac{\cosh^{N}\left(u-\eta\right)}{\cosh^{N}\left(u+\eta\right)}\quad{\rm and}\quad d_{u}=1.
\end{equation}
Under substitution of these expressions and of the expression for
$b\left(u\right)$ in Eq.~(\ref{eq:b(u), c(u)}), the constraint
in Eq.~(\ref{eq:Bethe_equations_ABA_v0}) reads as 
\begin{equation}
\frac{\cosh^{N}\left(u_{m}-\eta\right)}{\cosh^{N}\left(u_{m}+\eta\right)}=\prod_{l=1\neq m}^{M}\frac{\sinh\left(u_{m}-u_{l}-2\eta\right)}{\sinh\left(u_{m}-u_{l}+2\eta\right)}\label{eq:Bethe_equations_ABA}
\end{equation}
and the eigenvalue $\mathcal{T}_{u}$ in Eq.~(\ref{eq:Tau_u_v0})
as 
\begin{equation}
\mathcal{T}_{u}=\frac{\cosh^{N}\left(u-\eta\right)}{\cosh^{N}\left(u+\eta\right)}\prod_{m=1}^{M}\frac{\sinh\left(u-u_{m}+2\eta\right)}{\sinh\left(u-u_{m}\right)}+\prod_{m=1}^{M}\frac{\sinh\left(u_{m}-u+2\eta\right)}{\sinh\left(u_{m}-u\right)}.\label{eq:T_u}
\end{equation}

The set of equations~(\ref{eq:Bethe_equations_ABA}) are the Bethe
equations for the XXZ model in Orbach parametrisation \citep{Gaudin_book}.
Substitution of the inverse mapping from Orbach (which is also known
as algebraic) to the coordinate parametrisation,
\begin{align}
u_{m} & =\frac{1}{2}\ln\left(\frac{1-e^{iq_{m}-2\eta}}{1-e^{-iq_{m}-2\eta}}\right)-\frac{iq_{m}}{2},\label{eq:um_qm}\\
\eta & =\frac{1}{2}{\rm acosh}\Delta,\label{eq:eta_Delta}
\end{align}
into Eq.~(\ref{eq:Bethe_equations_ABA}) recovers the Bethe equations
in the coordinate representation in Eq.~(\ref{eq:LiebWu_Jm0}) with
the two-magnon scattering phase in Eq.~(\ref{eq:phi_lm_XXZ}).

The advantage of the algebraic over the coordinate representation
of the Bethe wave function can already be seen in calculation of the
scalar product of two Bethe states. In the algebraic representation
of the scalar product of two Bethe states $\left\langle \mathbf{v}\right|$
and $\left|\mathbf{u}\right\rangle $ in Eq.~(\ref{eq:Bethe_wavefunction_ABA}),
each $B\left(u_{j}\right)$ operator in the bra state can be commuted
through the product of all $C\left(u_{j}\right)$ operators in the
ket state using the commutation relations in Eqs.~(\ref{eq:BuCv}--\ref{eq:DuCv}).
At the end of the commutation procedure, the $B\left(u_{j}\right)$
operators acting upon the vacuum states $\left|\Downarrow\right\rangle $
give zero and the $A\left(u_{j}\right)$ and $D\left(u_{j}\right)$
operators generated by the $B\left(u_{j}\right)$ and $C\left(u_{j}\right)$
commutation relation (\ref{eq:BuCv}) give their vacuum eigenvalues
$a_{j}$ and $d_{j}$. When the ket state $\left|\mathbf{u}\right\rangle $
is parametrised by $u_{j}$ that satisfy the Bethe equations (\ref{eq:Bethe_equations_ABA}),
the result of such commutation for the scalar product of two Bethe
states can be written in a compact form as a determinant of an $M\times M$
matrix, which is also known as the Slavnov formula \citep{Slavnov89},
\begin{equation}
\left\langle \mathbf{v}|\mathbf{u}\right\rangle =\frac{\prod_{lm}\sinh\left(v_{l}-u_{m}\right)\det\hat{G}}{\prod_{l<m}\sinh\left(v_{l}-v_{m}\right)\prod_{l<m}\sinh\left(u_{l}-u_{m}\right)}\label{eq:Scalar_product_ABA} ,
\end{equation}
where the $M\times M$ matrix $\hat{G}$ is given by the derivatives
of the eigenvalue of the transition matrix $\mathcal{T}_{u}$ as $G_{ab}=\partial_{u_{a}}\mathcal{T}\left(v_{b}\right)$.
The explicit form of these derivatives for the $\mathcal{T}_{u}$
in Eq.~(\ref{eq:T_u}) is 
\begin{equation}
G_{ab}=\frac{\cosh^{N}\left(v_{b}-\eta\right)}{\cosh^{N}\left(v_{b}+\eta\right)}\frac{\sinh\left(2\eta\right)}{\sinh^{2}\left(v_{b}-u_{a}\right)}\prod_{l=1\neq a}^{M}\frac{\sinh\left(v_{b}-u_{l}+2\eta\right)}{\sinh\left(v_{b}-u_{l}\right)}-\frac{\sinh\left(2\eta\right)}{\sinh^{2}\left(u_{a}-v_{b}\right)}\prod_{l=1\neq a}^{M}\frac{\sinh\left(u_{l}-v_{b}+2\eta\right)}{\sinh\left(u_{l}-v_{b}\right)}.\label{eq:B_ab}
\end{equation}
When $v_{j}$ satisfy the Bethe equations (\ref{eq:Bethe_equations_ABA})
instead of $u_{j}$, the result is almost the same in Eq.~(\ref{eq:Scalar_product_ABA}),
with the only difference that the derivatives of the eigenvalue of
the transition matrix taken over the $v_{j}$ instead of $u_{j}$ as
$G_{ab}=\partial_{v_{a}}\mathcal{T}\left(u_{b}\right)$. 

\subsection{Normalisation factor}

The normalisation factor of the Bethe states in the algebraic representation
in Eq.~(\ref{eq:Bethe_wavefunction_ABA}) can be evaluated by applying
the formula for the scalar product in Eq.~(\ref{eq:Scalar_product_ABA})
on the pair of the same states, $Z^{2}=\left\langle \mathbf{u}|\mathbf{u}\right\rangle $.
Since some of the matrix elements in Eq.~(\ref{eq:B_ab}) are divergent
under the direct substitution of $\mathbf{v}=\mathbf{u}$, which is,
however, regularised by the zeros in the prefactor in Eq.~(\ref{eq:Scalar_product_ABA}),
this substitution needs to be evaluated by taking the $\mathbf{v}\rightarrow\mathbf{u}$
limit, which gives \citep{Korepin82}
\begin{equation}
Z^{2}=\sinh^{M}\left(2\eta\right)\prod_{l\neq m}\frac{\sinh\left(u_{l}-u_{m}+2\eta\right)}{\sinh\left(u_{l}-u_{m}\right)}\det\hat{F},\label{eq:Z2_ABA}
\end{equation}
where the matrix $\hat{F}$ is 
\begin{equation}
F_{ab}=\begin{cases}
-N\frac{\sinh\left(2\eta\right)}{\cosh\left(u_{a}+\eta\right)\cosh\left(u_{a}-\eta\right)}-\sum_{l=1\neq a}^{M}\frac{\sinh\left(4\eta\right)}{\sinh\left(u_{a}-u_{l}-2\eta\right)\sinh\left(u_{a}-u_{l}+2\eta\right)}, & a=b,\\
\frac{\sinh\left(4\eta\right)}{\sinh\left(u_{b}-u_{a}-2\eta\right)\sinh\left(u_{b}-u_{a}+2\eta\right)}, & a\neq b.
\end{cases}
\end{equation}

The mapping of the last result back into the coordinate representation,
in which we calculate the spin matrix element in this work, is done
by substituting the relations in Eqs.~(\ref{eq:um_qm}, \ref{eq:eta_Delta})
in Eq.~(\ref{eq:Z2_ABA}). In the Heisenberg limit, which we also
need for the Hubbard model in this work, this substitution becomes
degenerate since $\eta=0$ and $u_{m}=i\pi/2$ for $\Delta=1$. Therefore,
the $\eta\rightarrow0$ limit needs to be taken in doing this inverse mapping.
Expanding Eq.~(\ref{eq:um_qm}) upto the linear order in $\eta$,
\begin{equation}
u_{m}=\frac{i\pi}{2}+i\eta\cot\frac{q_{m}}{2},\label{eq:um_qm_Heisenberg_limit}
\end{equation}
substituting the last expansion in Eq.~(\ref{eq:Z2_ABA}), and taking
the $\eta\rightarrow0$ limit 
\begin{equation}
\lim_{\eta\rightarrow0}\left.Z^{2}\right|_{u_{m}\rightarrow\frac{i\pi}{2}+i\eta\cot\frac{q_{m}}{2}},\label{eq:Heisenberg_limit_ABA}
\end{equation}
gives
\begin{equation}
Z^{2}=\left(-4\right)^{M}\prod_{m}\sin^{2}\frac{q_{m}}{2}\prod_{l\neq m}\frac{\cot\frac{q_{m}}{2}-\cot\frac{q_{l}}{2}-2i}{\cot\frac{q_{m}}{2}-\cot\frac{q_{l}}{2}}\det\hat{Q}_{0}.\label{eq:Z2_ABA_coordinate}
\end{equation}
Here the determinant of the matrix $\hat{Q}_{0}$ given by Eq.~(\ref{eq:Q_ml0})
is the same as in the Gaudin normalisation factor in the coordinate
representation \citep{Gaudin81} but the prefactor in the algebraic
representation is not unity but a function of spin quasimomenta. Since
we calculate the spin matrix element in the coordinate representation
using the algebraic method in this work, the non-unity prefactor in
Eq.~(\ref{eq:Z2_ABA_coordinate}) will be taken into account where
needed in what follows.

\section{$t/U$ expansion of the matrix element $\langle f|c_{1\uparrow}^{\pm}|0\rangle$}

The matrix elements of the ladder operators $c_{1\alpha}^{\pm}$ are
needed for calculation of the Green function and of a range of observables
relevant in practice. Since the Hubbard model without a magnetic field
in Eq.~(1) of the main text has the symmetry w.r.t. $\uparrow\leftrightarrow\downarrow$,
it is sufficient to consider only $\alpha=\uparrow$. Let us start
from the annihilation operators $c_{k\uparrow}$. Its matrix element
that is required for the Green function at zero temperature is evaluated
between the ground state $0$ and an excited state $f$, $\langle f|c_{1\uparrow}|0\rangle$.
For large $U$, the expansion of the wave function in Eq.~(\ref{eq:Psi01})
gives the $t/U$ expansion of the matrix element as 
\begin{equation}
\langle f|c_{1\uparrow}|0\rangle=\langle f|c_{1\uparrow}|0\rangle^{0}+\frac{t}{U}\langle f|c_{1\uparrow}|0\rangle^{1}.\label{eq:MEminus}
\end{equation}

The zeroth term for the matrix element here factorises into the charge
and spin parts,
\begin{equation}
\langle f|c_{1\uparrow}|0\rangle^{0}=\langle f|c_{1\uparrow}|0\rangle_{\rm c}^{0}\cdot\langle f|c_{1\uparrow}|0\rangle_{\rm s}^{0},\label{eq:MEminus0}
\end{equation}
since the zeroth order of the wave function in Eq.~(\ref{eq:Psi0})
is factorised into these two sectors. The charge part is evaluated
as an $N$-fold sum over the charge coordinates $\mathbf{j}$ using
Slater determinants in Eq.~(\ref{eq:psi_j0c}), producing a determinant
of the Vandermonde type as a result \citep{Ogata90},
\begin{equation}
\langle f|c_{1\uparrow}|0\rangle_{\rm c}^{0}=\frac{1}{L^{-N+\frac{1}{2}}}\det\hat{C}_{0},\label{eq:ME_c0}
\end{equation}
where the matrix elements of the $N\times N$ matrix $\hat{C}_{0}$
are 
\begin{equation}
C_{ab}^{0}=\begin{cases}
e^{-i\frac{k_{a}^{00}\left(L-1\right)}{2}}, & b=1,\\
\frac{\sin\left(\frac{k_{a}^{00}-k_{b-1}^{f0}}{2}L\right)}{\sin\left(\frac{k_{a}^{00}-k_{b-1}^{f0}}{2}\right)}, & b>1.
\end{cases}\label{eq:C_ab0}
\end{equation}
Substitution of the explicit solutions for the charge quasimomenta
in the $U=\infty$ limit from Eq.~(\ref{eq:k_j0}) in the above gives
\begin{equation}
C_{ab}^{0}=\begin{cases}
1, & b=1,\\
2\frac{\sin\big(\frac{P_{\rm s}^{0}-P_{\rm s}^{f}}{2}\big)}{k_{a}^{00}-k_{b-1}^{f0}}, & b>1,
\end{cases}\label{eq:C_ab0_low_density}
\end{equation}
where the limit of low-particle density, $N/L\ll1$, was also taken
and the overall real phase was neglected since it does not affect the
observables. The last expression is presented in the main text after
Eq.~(4).

The spin part of the matrix element in Eq.~(\ref{eq:MEminus0}) is
evaluated by means of the algebraic Bethe ansatz \citep{Sklyanin79}.
This technique was successfully applied to calculation of the correlation
function of the antiferromagnetic 1D-Heisenberg model \citep{Kitanine99,Kitanine00}.
However, this result cannot be used for the Hubbard model directly,
since the ladder operator $c_{1\uparrow}$ here changes the length
of the spin chain by one site, making the construction of \citep{Sklyanin79}
for the bra $\langle f_s^0|$ and ket $|0_s^0\rangle$ states in the spin part of Eq.~(\ref{eq:MEminus0})
incompatible with each other. This problem was resolved by developing
a representation of the algebra for the longer spin chain through
the other algebra for the shorter one in \citep{Tsyplyatyev22}, where
the spin part of the matrix element for the Hubbard model was obtained
as
\begin{equation}
\langle f|c_{1\uparrow}|0\rangle_{\rm s}^{0}=\frac{1}{Z_{0}Z_{f}}\frac{\prod_{lm}\left(e^{iq_{l}^{f0}}+e^{-iq_{m}^{00}}-2\right)\det\hat{R}_{00}}{\prod_{l\neq m}\left(e^{iq_{l}^{f0}}+e^{-iq_{m}^{f0}}-2\right)^{\frac{1}{2}}\prod_{l\neq m}\left(e^{iq_{l}^{00}}+e^{-iq_{m}^{00}}-2\right)^{\frac{1}{2}}},\label{eq:ME_s0}
\end{equation}
where the matrix elements of the $M\times M$ matrix $\hat{R}_{00}$
are 
\begin{equation}
R_{ab}^{00}=\begin{cases}
\frac{e^{i\left(N-1\right)q_{b}^{00}}\prod_{l\neq a}\frac{e^{iq_{l}^{f0}+iq_{b}^{00}}+1-2e^{iq_{l}^{f0}}}{2e^{iq_{b}^{00}}-e^{iq_{l}^{f0}+iq_{b}^{00}}-1}-1}{\left(e^{-q_{a}^{f0}}-e^{-iq_{b}^{00}}\right)\left(e^{q_{a}^{f0}}-e^{-iq_{b}^{00}}-2\right)}, & a<M,\\
e^{ik_{b}^{00}}\frac{\prod_{l\neq b}\big(e^{iq_{l}^{00}}+e^{-iq_{b}^{00}}-2\big)}{\prod_{l}\big(e^{iq_{l}^{f0}}+e^{-iq_{b}^{00}}-2\big)}, & a=M,
\end{cases}\label{eq:R_ab00}
\end{equation}
and the normalisation factors $Z_{0/f}=\sqrt{{\rm det}\hat{Q}_{0}^{f/0}}$
are given by the determinant of the matrix $\hat{Q}_{0}^{f/0}$ in
Eq.~(\ref{eq:Q_ml0}) with the spin quasimomenta in the $U=\infty$
limit for the bra $\langle f_s^0|$ and the ket $|0_s^0\rangle$ states, respectively. 

The linear term in Eq.~(\ref{eq:MEminus}) separates into three independent
terms as
\begin{equation}
\langle f|c_{1\uparrow}|0\rangle^{1}=\langle f|c_{1\uparrow}|0\rangle_{\rm c}^{1}+\langle f|c_{1\uparrow}|0\rangle_{\rm cs}^{1}+\langle f|c_{1\uparrow}|0\rangle_{\rm s}^{1}\label{eq:MEminus1}
\end{equation}
following separation of linear term of the wave function in Eq.~(\ref{eq:Psi1})
into the same three terms. We will evaluate each of these contributions
to the matrix element in the linear order in Eq.~(\ref{eq:MEminus1})
independently below.

\subsection{Charge part}

The first term in Eq.~(\ref{eq:MEminus1}) has two contributions
originating from the linear terms in the bra and ket wave functions
of the charge type in Eq.~(\ref{eq:Psi1c}),
\begin{equation}
\langle f|c_{1\uparrow}|0\rangle_{\rm c}^{1}=\left(\langle f_{\rm c}^{0}|c_{1\uparrow}|0_{\rm c}^{1}\rangle+\langle f_{\rm c}^{1}|c_{1\uparrow}|0_{\rm c}^{0}\rangle\right)\langle f|c_{1\uparrow}|0\rangle_{\rm s}^{0},\label{eq:ME_c1def}
\end{equation}
where the spin part in the zeroth order in $t/U$ is already given
by Eq.~(\ref{eq:ME_s0}). We simplify evaluation of the $N$-fold
sum over the coordinates in the charge part by noting that the linear
term of the wave function $\left|\Psi_{\rm c}^{1}\right\rangle $ in Eq.~(\ref{eq:Psi1c})
can be obtained from the zeroth order term $\left|\Psi_{\rm c}^{0}\right\rangle $
in Eq.~(\ref{eq:Psi_c0}) by substituting $\mathbf{k}^{0}\rightarrow\mathbf{k}^{0}+g\mathbf{k}^{1}$
in the latter, taking the derivative of the result w.r.t.\ $g$, and
by taking the $g\rightarrow0$ limit at the end,
\begin{equation}
\left|\Psi_{\rm c}^{1}\right\rangle =\text{\ensuremath{\lim}}_{g\rightarrow0}d_{g}\left(\left|\Psi_{\rm c}^{0}\right\rangle _{\mathbf{k}^{0}\rightarrow\mathbf{k}^{0}+g\mathbf{k}^{1}}\right) ,
\end{equation}
where the charge quasimomenta in the first $t/U$ order $\mathbf{k}^{1}$
are given by Eq.~(\ref{eq:k_j1}).

Using this trick for both terms in Eq.~(\ref{eq:ME_c1def}) as $\mathbf{k}_{0/f}^{0}\rightarrow\mathbf{k}_{0/f}^{0}+g\mathbf{k}_{0/f}^{1}$
and commuting the $\text{\ensuremath{\lim}}_{g\rightarrow0}d_{g}$
with the $N$-fold sum over the charge coordinates $\mathbf{j}$,
we express the whole charge as a single term,
\begin{equation}
\langle f|c_{1\uparrow}|0\rangle_{\rm c}^{1}=\langle f|c_{1\uparrow}|0\rangle_{\rm s}^{0}\lim_{g\rightarrow0}d_{g}\langle f_{\rm c}^{0}+gf_{\rm c}^{1}|c_{1\uparrow}|0_{\rm c}^{0}+g0_{\rm c}^{1}\rangle^{0},\label{eq:ME_c1_v2}
\end{equation}
where application of the chain rule in calculating the derivative
$d_{g}$ recovers the two contributions in Eq.~(\ref{eq:ME_c1def}).
Calculation of the charge matrix element for the zeroth-order wave
functions $\left|\Psi_{\rm c}^{0}\right\rangle $ in the last expression
was already done in \citep{Ogata90}. Additional caution needs to be
used to reuse this calculation here, since the $g\rightarrow0$ limit
and the $d_{g}$ derivative in the first order in Eq.~(\ref{eq:ME_c1_v2})
do not commute with each other and also do not commute with the low-density
limit taken in Eq.~(\ref{eq:C_ab0_low_density}). Therefore, we use
the result in Eq.~(\ref{eq:C_ab0}), substitute $\mathbf{k}_{0/f}^{0}\rightarrow\mathbf{k}_{0/f}^{0}+g\mathbf{k}_{0/f}^{1}$
into it, and expand the resulting matrix under the determinant up to linear order in $g$ as 
\begin{equation}
\hat{C_{0}}+g\hat{C}_{1}\label{eq:C0+gC1} ,
\end{equation}
where the matrix elements in the first order in $g$ are 
\begin{equation}
C_{ab}^{1}=\begin{cases}
-i\frac{k_{a}^{01}\big(L-1\big)}{2}, & b=1,\\
2\Bigg(\frac{L\cos\frac{P_{\rm s}^{0}-P_{\rm s}^{f}}{2}}{2}-\frac{\sin\frac{P_{\rm s}^{0}-P_{\rm s}^{f}}{2}}{k_{a}^{00}-k_{b-1}^{f0}}\Bigg)\frac{k_{a}^{01}-k_{b-1}^{f1}}{k_{a}^{00}-k_{b-1}^{f0}}, & b>1,
\end{cases}\label{eq:C_ab1}
\end{equation}
in which the low-density limit was taken at the last step and the
matrix element of $\hat{C}_{0}$ in the low-density limit are given
in Eq.~(\ref{eq:C_ab0_low_density}). Then, we use the Jacobi formula
for evaluating the derivative w.r.t. $g$ of the
determinant from Eq.~(\ref{eq:ME_c0}) with the matrix (\ref{eq:C0+gC1})
under it as 
\begin{equation}
d_{g}\det\left(\hat{C_{0}}+g\hat{C}_{1}\right)=\det\left(\hat{C_{0}}+g\hat{C}_{1}\right){\rm tr}\left[\left(\hat{C_{0}}+g\hat{C}_{1}\right)^{-1}\hat{C}_{1}\right].\label{eq:d_gdetC0+gC1}
\end{equation}

Lastly, substitution of Eq.~(\ref{eq:ME_c0}) with the last expression
into Eq.~(\ref{eq:ME_c1_v2}) gives 
\begin{equation}
\langle f|c_{1\uparrow}|0\rangle_{\rm c}^{1}=\langle f|c_{1\uparrow}|0\rangle_{\rm s}^{0}\frac{1}{L^{-N+\frac{1}{2}}}\det\hat{C_{0}}{\rm tr}\left(\hat{C_{0}}^{-1}\hat{C}_{1}\right),\label{eq:ME_c1_v3}
\end{equation}
where the $g\rightarrow0$ limit also insures that all the higher-than-linear
terms in the $g$-expansion in Eq.~(\ref{eq:C0+gC1}) do not contribute
to the charge part of the matrix element. Identification of the charge
part of the matrix element in the zeroth order $\langle f|c_{1\uparrow}|0\rangle_{\rm c}^{0}$
in Eq.~(\ref{eq:ME_c1_v3}) by means of Eq.~(\ref{eq:ME_c0}) allows
to factor out the whole zeroth-order matrix element in Eq.~(\ref{eq:ME_c1_v3})
as 
\begin{equation}
\langle f|c_{1\uparrow}|0\rangle_{\rm c}^{1}=\langle f|c_{1\uparrow}|0\rangle^{0}T_{\rm c},
\end{equation}
where 
\begin{equation}
T_{\rm c}={\rm tr}\left(\hat{C_{0}}^{-1}\hat{C}_{1}\right).\label{eq:T_c}
\end{equation}
The results in Eqs.~(\ref{eq:C_ab1}, \ref{eq:T_c}) are presented
in Eqs.~(5--7) of the main paper. 

\subsection{Spin part}

The third term in Eq.~(\ref{eq:MEminus1}), the spin contribution
to the linear term, can be evaluated using the same trick with the
derivative and the limit, as for the charge term in the first order
in Eq.~(\ref{eq:ME_c1_v2}), since the linear term of the wave function
of the spin type $\left|\Psi_{\rm s}^{1}\right\rangle $ in Eq.~(\ref{eq:Psi1})
can also be expressed using the zeroth-order term $\left|\Psi_{\rm s}^{0}\right\rangle $
in Eq.~(\ref{eq:Psi_s0}) as
\begin{equation}
\left|\Psi_{\rm s}^{1}\right\rangle =\text{\ensuremath{\lim}}_{g\rightarrow0}d_{g}\left(\left|\Psi_{\rm s}^{0}\right\rangle _{\mathbf{q}^{0}\rightarrow\mathbf{q}^{0}+g\mathbf{q}^{1}}\right)\label{eq:Psi_s1_dPsi_s0} ,
\end{equation}
where the spin quasimomenta in the first $t/U$ order $\mathbf{q}^{1}$
are given by Eq.~(\ref{eq:q1m}). However, the application of this
trick to the spin part of matrix element is somewhat less straightforward.
In the charge part in Eq.~(\ref{eq:ME_c1def}), the $N$-fold sum
over the charge coordinates is performed in the first-quantisation
formalism, which imposes no additional requirements on the bra and
ket states so that we can use the trick for both of them simultaneously
in Eq.~(\ref{eq:ME_c1_v2}). For the spin part, the $M$-fold sum
over the spin coordinates is performed indirectly using the algebraic
Bethe ansatz by means of the Slavnov formula \citep{Slavnov89}
in Eq.~(\ref{eq:Scalar_product_ABA}). Application of this formula
requires one of the bra or ket states to be an eigenstate, which is
not satisfied when both of the states are shifted by $g\mathbf{q}_{0/f}^{1}$
simultaneously. Thus, we have to apply the trick for the bra and ket
states in the spin part of the matrix element separately. 

Moreover, the spin part of the matrix element in the zeroth order
in Eq.~(\ref{eq:ME_s0}) was derived under the specific condition
of the bra state being an eigenstate in \citep{Tsyplyatyev22}, so we have
to derive additionally the explicit expression for the spin matrix
element in the zeroth order under the condition of ket state being
an eigenstate here. We will do so by repeating the steps from \citep{Tsyplyatyev22},
but with the different assumption of the ket instead of the bra state
being an eigenstate. Our starting point is the representation of the
spin part of the matrix element in the zeroth order in the representation
of the algebraic Bethe ansatz,
\begin{equation}
\langle f|c_{1\uparrow}|0\rangle_{\rm s}^{0}=e^{-i\left(P_{\rm s}^{0}-P_{\rm s}^{f}\right)\left(N-1\right)}\frac{\langle\mathbf{v}|D_{N}S_{N}^{-}|\mathbf{u}\rangle}{Z_{f}Z_{0}},\label{eq:ME_s0_v2_def}
\end{equation}
where the decomposition of the electron ladder operator into the spin
and charge parts in Eq.~(\ref{eq:c_jup}) was used, $\mathbf{v}$
are the spin quasimomenta $\mathbf{q}_{0}^{f}$ and $\mathbf{u}$
are the spin quasimomenta $\mathbf{q}_{0}^{0}$ in the algebraic representation
Eq.~(\ref{eq:um_qm}), $Z_{f/0}$ are the normalisation factors of
the Bethe functions in the algebraic representation in Eq.~(\ref{eq:Z2_ABA}),
and the spin chain was shifted $N-1$ times to the right in order
to simplify the algebraic manipulations in what follows. Here we assume
that $|\mathbf{u}\rangle$ is an eigenstate instead of $\langle\mathbf{v}|$. 

First, we express the local spin operator $S_{N}^{-}$ in Eq.~(\ref{eq:ME_s0_v2_def})
in terms of the algebraic Bethe-ansatz operators in Eq.~(\ref{eq:T_matrix})
by means of the Drinfeld twist \citep{Drinfeld83} as 
\begin{equation}
S_{N}^{-}=B_{\xi}\tau_{\xi}^{N-1},
\end{equation}
where $\xi=-i\pi/2+\eta$ and the transition matrix $\tau_{\xi}$
is given by Eq.~(\ref{eq:T_u}). Substitution of the last expression
into the matrix element in the r.h.s.\ of Eq.~(\ref{eq:ME_s0_v2_def})
gives 
\begin{equation}
\langle\mathbf{v}|D_{N}S_{N}^{-}|\mathbf{u}\rangle=e^{iP_{\rm s}^{0}\left(N-1\right)}\langle\Downarrow|\prod_{m=1}^{M-1}B^{N-1}\left(v_{m}\right)B_{\xi}^{N}\prod_{m=1}^{M}C^{N}\left(u_{m}\right)|\Downarrow\rangle,\label{eq:ME_s0_ABA_f}
\end{equation}
where $C^{N}\left(u\right)$ and $B^{N}\left(u\right)$ are the Bethe-ansatz operators constructed for the chain of $N$ spins and $B^{N-1}\left(u\right)$
are the Bethe-ansatz operators for the chain of $N-1$ spins due to
the $D_{N}$ operator in Eq.~(\ref{eq:ME_s0_v2_def}).

The last expression is almost a scalar product of two Bethe states,
but the algebraic Bethe ansatz operators in the bra and ket states
are constructed for the chains of different lengths, making the Slavnov
formula in Eq.~(\ref{eq:Scalar_product_ABA}) inapplicable. In order
to restore its applicability, we need to express the operators in
the bra state using the same algebra as in the ket state by using
the construction of the Bethe-ansatz operators in Eq.~(\ref{eq:T_XXZ}).
Singling out the operators for the chain of $N-1$ spins in this construction
for $N$ spins we obtain 
\begin{equation}
\left(\begin{array}{cc}
A_{u}^{N} & B_{u}^{N}\\
C_{u}^{N} & D_{u}^{N}
\end{array}\right)=\left(\begin{array}{cc}
A_{u}^{N-1} & B_{u}^{N-1}\\
C_{u}^{N-1} & D_{u}^{N-1}
\end{array}\right)\left(\begin{array}{cc}
\frac{\cosh\left(u+2\eta S_{N}^{z}\right)}{\cosh\left(u+\eta\right)} & -i\frac{\sinh2\eta S_{N}^{-}}{\cosh\left(u+\eta\right)}\\
-i\frac{\sinh2\eta S_{N}^{+}}{\cosh\left(u+\eta\right)} & \frac{\cosh\left(u-2\eta S_{N}^{z}\right)}{\cosh\left(u+\eta\right)}
\end{array}\right).
\end{equation}
Multiplying this equation by the inverse of the matrix in the second
factor in the r.h.s., and picking the top-right element of the resulting
matrix equation, we get
\begin{equation}
B_{u}^{N-1}=\frac{\cosh\left(u+\eta\right)}{\cosh\left(u-2\eta S_{N}^{z}\right)}B_{u}^{N}+i\frac{\sinh2\eta}{\cosh\left(u-2\eta S_{N}^{z}\right)}S_{N}^{-}A_{u}^{N-1}.\label{eq:B_uN-1}
\end{equation}
Lastly, we substitute this expression for $B_{u}^{N-1}$ into Eq.~(\ref{eq:ME_s0_ABA_f})
and obtain 
\begin{equation}
\langle\mathbf{v}|D_{N}S_{N}^{-}|\mathbf{u}\rangle=e^{iP_{\rm s}^{0}\left(N-1\right)}\left\langle \Downarrow\right|\prod_{m=1}^{M-1}B^{N}\left(v_{j}\right)B^{N}\left(\xi\right)\prod_{m=1}^{M}C^{N}\left(u_{j}\right)\left|\Downarrow\right\rangle ,\label{eq:ME_s0_ABA_f_v2}
\end{equation}
where the second term in Eq.~(\ref{eq:B_uN-1}) does not contribute
since $S_{N}^{-}B_{\xi}^{N}\sim S_{N}^{-}S_{N}^{-}=0$ and 
\begin{equation}
\frac{\cosh\left(v_{m}+\eta\right)}{\cosh\left(v_{m}-2\eta S_{N}^{z}\right)}B_{\xi}^{N}=B_{\xi}^{N} ,
\end{equation}
 since the $N^{{\rm th}}$ spin to the left of the $B_{\xi}^{N}$
operator is always in the $\downarrow$ state.

The expression in Eq.~(\ref{eq:ME_s0_ABA_f_v2}) is a scalar product
of two Bethe states constructed for the same chain consisting of $N$
spins. Applying the Slavnov formula in Eq.~(\ref{eq:Scalar_product_ABA})
to it, performing the inverse mapping to the coordinate representation
in Eq.~(\ref{eq:um_qm_Heisenberg_limit}), and taking the $\eta\rightarrow0$
limit of the result, as in Eq.~(\ref{eq:Heisenberg_limit_ABA}),
we obtain 
\begin{equation}
\langle\mathbf{v}|D_{N}S_{N}^{-}|\mathbf{u}\rangle=e^{iP_{\rm s}^{0}\left(N-1\right)}\frac{2^{M}i^{M}\prod_{l}\left(1+i\cot\frac{q_{l}^{00}}{2}\right)\prod_{lm}\left(\cot\frac{q_{m}^{00}}{2}-\cot\frac{q_{l}^{f0}}{2}\right)\det\hat{G}_{f}}{\prod_{l}\left(1-i\cot\frac{q_{l}^{f0}}{2}\right)\prod_{l<m}\left(\cot\frac{q_{l}^{f0}}{2}-\cot\frac{q_{m}^{f0}}{2}\right)\prod_{l<m}\left(\cot\frac{q_{l}^{00}}{2}-\cot\frac{q_{m}^{00}}{2}\right)},\label{eq:ME_s0_ABA_f_v3}
\end{equation}
where the matrix $\hat{G_{f}}$ is 
\begin{equation}
G_{ab}^{f}=\begin{cases}
\frac{e^{iq_{b}^{f0}N}\prod_{l=1\neq a}^{M}\frac{\cot\frac{q_{b}^{f0}}{2}-\cot\frac{q_{l}^{00}}{2}-2i}{\cot\frac{q_{b}^{f0}}{2}-\cot\frac{q_{l}^{00}}{2}}-\prod_{l=1\neq a}^{M}\frac{\cot\frac{q_{b}^{f0}}{2}-\cot\frac{q_{l}^{00}}{2}+2i}{\cot\frac{q_{b}^{f0}}{2}-\cot\frac{q_{l}^{00}}{2}}}{\left(\cot\frac{q_{b}^{f0}}{2}-\cot\frac{q_{a}^{00}}{2}\right)^{2}}, & b<M,\\
\frac{1}{\left(1-i\cot\frac{q_{a}^{00}}{2}\right)\left(1+i\cot\frac{q_{a}^{00}}{2}\right)}, & b=M.
\end{cases}
\end{equation}

Substituting the result in Eq.~(\ref{eq:ME_s0_ABA_f_v3}) and the
normalisation of the algebraic Bethe states in Eq.~(\ref{eq:Z2_ABA_coordinate})
into Eq.~(\ref{eq:ME_s0_v2_def}) and rearranging terms, we obtain
the same expression for the spin matrix element as in Eq.~(\ref{eq:ME_s0}),
\begin{equation}
\langle f|c_{1\uparrow}|0\rangle_{\rm s}^{0}=\frac{1}{Z_{0}Z_{f}}\frac{\prod_{lm}\left(e^{iq_{l}^{f0}}+e^{-iq_{m}^{00}}-2\right)\det\hat{R}_{f0}}{\prod_{l\neq m}\left(e^{iq_{l}^{f0}}+e^{-e^{iq_{m}^{f0}}}-2\right)^{\frac{1}{2}}\prod_{l\neq m}\left(e^{iq_{l}^{00}}+e^{-iq_{m}^{00}}-2\right)^{\frac{1}{2}}},\label{eq:ME_s0_v2}
\end{equation}
but with a different matrix $\hat{R}_{f0}$ under the determinant, the elements of which are 
\begin{equation}
R_{ab}^{f0}=\begin{cases}
\frac{e^{iq_{b}^{f0}N}\prod_{l=1\neq a}^{M}\left(-\frac{e^{iq_{b}^{f0}+iq_{l}^{00}}+1-2e^{iq_{l}^{00}}}{e^{iq_{b}^{f0}+iq_{l}^{00}}+1-2e^{iq_{b}^{f0}}}\right)-1}{\left(e^{-iq_{b}^{f0}}-e^{-iq_{a}^{00}}\right)\left(e^{iq_{a}^{00}}+e^{-iq_{b}^{f0}}-2\right)}, & b<M,\\
1, & b=M,
\end{cases}
\end{equation}
instead of the matrix $\hat{R}_{00}$ in Eq.~(\ref{eq:R_ab00}).
In the result in Eq.~(\ref{eq:ME_s0_v2}), the normalisation factors
$Z_{0/f}=\sqrt{{\rm det}\hat{Q}_{0}^{f/0}}$ are in the coordinate
representation and the property $P_{\rm s}^{f}\left(N-1\right)=2\pi\times{\rm integer\ number}$
so that $\exp\left(iP_{\rm s}^{f}\left(N-1\right)\right)=1$ was used.
In the $U=\infty$ limit, only one spin matrix in Eq.~(\ref{eq:ME_s0})
was needed for the observables in \citep{Tsyplyatyev22} since $|\det\hat{R}_{f0}|^{2}=|\det\hat{R}_{00}|^{2}$,
but in the first $t/U$-order the relative phase between $\det\hat{R}_{00}$
and $\det\hat{R}_{f0}$ affects the result through derivatives, so
both have to be introduced and evaluated. 

Turning back to the spin part of the matrix element in the first order,
we are now ready to evaluate the third term in Eq.~(\ref{eq:MEminus1}),
which has two contributions originating from the linear terms in the
bra and ket wave functions of the spin type in Eq.~(\ref{eq:Psi1s}),
\begin{equation}
\langle f|c_{1\uparrow}|0\rangle_{\rm s}^{1}=\langle f|c_{1\uparrow}|0\rangle_{\rm c}^{0}\left(\langle f_{\rm s}^{0}|c_{1\uparrow}|0_{\rm s}^{1}\rangle+\langle f_{\rm s}^{1}|c_{1\uparrow}|0_{\rm s}^{0}\rangle\right),\label{eq:ME_s0def}
\end{equation}
where the charge part in the zeroth $t/U$ order $\langle f|c_{1\uparrow}|0\rangle_{\rm c}^{0}$
is already given by Eq.~(\ref{eq:ME_c0}). Using the expression for
the wave function of the spin type in the first $t/U$ order in Eq.~(\ref{eq:Psi_s1_dPsi_s0})
and applying the same trick as for the charge part of the matrix element
in the linear order in Eq.~(\ref{eq:ME_c1_v2}) but only to the ket
state in first term in Eq.~(\ref{eq:ME_s0def}), we obtain 
\begin{equation}
\langle f_{\rm s}^{0}|c_{1\uparrow}|0_{\rm s}^{1}\rangle=\lim_{g\rightarrow0}d_{g}\langle f_{\rm s}^{0}|c_{1\uparrow}|0_{\rm s}^{0}+g0_{\rm s}^{1}\rangle^{0}.\label{eq:ME_s1_part_1_def}
\end{equation}
In the matrix element on the r.h.s.\ of the above expression, the spin
quasimomenta $\mathbf{q}_{0}^{f}$ in the bra state satisfy the Bethe
equations so evaluation of this spin part of the matrix element in
the zeroth $t/U$ order by means of the algebraic Bethe ansatz gives
Eq.~(\ref{eq:ME_s0}), in which the spin quasimomenta of the ground
state are shifted as $\mathbf{q}_{0}^{0}\rightarrow\mathbf{q}_{0}^{0}+g\mathbf{q}_{1}^{0}$.
Expanding the matrix under the determinant in Eq.~(\ref{eq:ME_s0})
after the shift up to linear order in $g$, similarly to Eq.~(\ref{eq:C0+gC1}),
we obtain 
\begin{equation}
\hat{R}_{00}+g\hat{R}_{01},\label{eq:R_00+gR_01} ,
\end{equation}
where the matrix in the first order in $g$ is $R_{ab}^{01}=\sum_{m}q_{m}^{01}\partial_{q_{m}^{00}}R_{ab}^{00}$.
Then, using the Jacobi formula for the derivative of a determinant
as in Eq.~(\ref{eq:ME_c1_v3}), we obtain 
\begin{multline}
\langle f_{\rm s}^{0}|c_{1\uparrow}|0_{\rm s}^{1}\rangle=\frac{1}{Z_{0}Z_{f}}\frac{\prod_{lm}\left(e^{iq_{l}^{f0}}+e^{-iq_{m}^{00}}-2\right)}{\prod_{l\neq m}\left(e^{iq_{l}^{f0}}+e^{-e^{iq_{m}^{f0}}}-2\right)^{\frac{1}{2}}\prod_{l\neq m}\left(e^{iq_{l}^{00}}+e^{-iq_{m}^{00}}-2\right)^{\frac{1}{2}}}\\
\times\left[\det\hat{R}_{00}{\rm tr}\left(\hat{R}_{00}^{-1}\hat{R}_{01}\right)-\frac{i}{2}\sum_{l\neq m}\frac{q_{l}^{01}e^{iq_{l}^{f0}}-q_{m}^{01}e^{-iq_{m}^{00}}}{e^{iq_{l}^{00}}+e^{-iq_{m}^{00}}-2}-i\sum_{lm}\frac{q_{m}^{01}e^{-iq_{m}^{00}}}{e^{iq_{l}^{f0}}+e^{-iq_{m}^{00}}-2}\right],\label{eq:ME_s1_part_1}
\end{multline}
where the last two terms containing the sums over the spin quasimomenta
in the first $t/U$ order $q_{m}^{01}$ originate from the derivatives
w.r.t.\ $g$ of the prefactor in front of the determinant in Eq.~(\ref{eq:ME_s0})
under the $\mathbf{q}_{0}^{0}\rightarrow\mathbf{q}_{0}^{0}+g\mathbf{q}_{1}^{0}$
shift, and the contribution from the $\mathbf{q}_{0}^{0}\rightarrow\mathbf{q}_{0}^{0}+g\mathbf{q}_{1}^{0}$
shift in the normalisation factor $Z_{0}$ is not taken into account
here. The latter involves additional algebraic complications and will
be considered separately as a part of Taylor expansion in $t/U$ for
the normalisation factor of the whole Lieb-Wu wave function in a separate
section below.

The second term in Eq.~(\ref{eq:ME_s0def}) can be expressed in the
same way, using the tricks in Eqs.~(\ref{eq:Psi_s1_dPsi_s0}) and
(\ref{eq:ME_c1_v2}) as
\begin{equation}
\langle f_{\rm s}^{0}|c_{1\uparrow}|0_{\rm s}^{1}\rangle=\lim_{g\rightarrow0}d_{g}\langle f_{\rm s}^{0}+gf_{\rm s}^{1}|c_{1\uparrow}|0_{\rm s}^{0}\rangle^{0}.\label{eq:ME_s1_part_2_def}
\end{equation}
Then, we repeat the same steps as after Eq.~(\ref{eq:ME_s1_part_1_def}),
but using the result for the sum over the spin coordinates in Eq.~(\ref{eq:ME_s0_v2})
instead of Eq.~(\ref{eq:ME_s0}) and obtain
\begin{multline}
\langle f_{\rm s}^{1}|c_{1\uparrow}|0_{\rm s}^{0}\rangle=\frac{1}{Z_{0}Z_{f}}\frac{\prod_{lm}\left(e^{iq_{l}^{f0}}+e^{-iq_{m}^{00}}-2\right)}{\prod_{l\neq m}\left(e^{iq_{l}^{f0}}+e^{-e^{iq_{m}^{f0}}}-2\right)^{\frac{1}{2}}\prod_{l\neq m}\left(e^{iq_{l}^{00}}+e^{-iq_{m}^{00}}-2\right)^{\frac{1}{2}}}\\
\times\left[\det\hat{R}_{f0}{\rm tr}\left(\hat{R}_{f0}^{-1}\hat{R}_{f1}\right)-\frac{i}{2}\sum_{l\neq m}\frac{q_{l}^{f1}e^{iq_{l}^{f0}}-q_{m}^{f1}e^{-iq_{m}^{f0}}}{e^{iq_{l}^{f0}}+e^{-iq_{m}^{f0}}-2}+i\sum_{lm}\frac{q_{l}^{f1}e^{iq_{l}^{f0}}}{e^{iq_{,}^{f0}}+e^{-iq_{m}^{00}}-2}\right],\label{eq:ME_s1_part_2}
\end{multline}
where $R_{ab}^{f1}=\sum_{m}q_{m}^{f1}\partial_{q_{m}^{f0}}R_{ab}^{f0}$
is the linear  term in the Taylor expansion of the matrix $\hat{R}_{f0}$ in $g$ under the
$\mathbf{q}_{0}^{f}\rightarrow\mathbf{q}_{0}^{f}+g\mathbf{q}_{1}^{f}$
shift of the spin quasimomenta and the last two
terms in the above expression containing the sums over the spin quasimomenta
in the first $t/U$ order $q_{m}^{f1}$ originate from the derivatives
w.r.t. $g$ of the prefactor under the same $\mathbf{q}_{0}^{f}\rightarrow\mathbf{q}_{0}^{f}+g\mathbf{q}_{1}^{f}$
shift.

Under the substitution of both terms in Eqs.~(\ref{eq:ME_s1_part_1})
and (\ref{eq:ME_s1_part_2}) into Eq.~(\ref{eq:ME_s0def}), the whole
zeroth-order matrix element can be identified as a factor using Eqs.~(\ref{eq:ME_s0}, \ref{eq:ME_s0_v2}),
and we find the spin contribution to the linear term as 
\begin{equation}
\langle f|c_{1\uparrow}|0\rangle_{\rm s}^{1}=\langle f|c_{1\uparrow}|0\rangle^{0}T_{\rm s},
\end{equation}
where 
\begin{multline}
T_{\rm s}=i\sum_{lm}\frac{q_{l}^{f1}e^{iq_{l}^{f0}}-q_{m}^{01}e^{-iq_{m}^{00}}}{e^{iq_{l}^{f0}}+e^{-iq_{m}^{00}}-2}-\frac{i}{2}\sum_{l\neq m}\frac{q_{l}^{01}e^{iq_{l}^{00}}-q_{m}^{01}e^{-iq_{m}^{00}}}{e^{iq_{l}^{00}}+e^{-iq_{m}^{00}}-2}+{\rm Tr}\big(\hat{R}_{00}^{-1}\hat{R}_{01}\big)\\
-\frac{i}{2}\sum_{l\neq m}\frac{q_{l}^{f1}e^{iq_{l}^{f0}}-q_{m}^{f1}e^{-iq_{m}^{f0}}}{e^{iq_{l}^{f0}}+e^{-iq_{m}^{f0}}-2}+{\rm Tr}\big(\hat{R}_{f0}^{-1}\hat{R}_{f1}\big).\label{eq:Ts}
\end{multline}
The last result is presented in Eq.~(8) of the main text.

\subsection{Mixing part}

In the second term in Eq.~(\ref{eq:MEminus1}), the sums over the
spin and charge coordinates are mixed with each other. Substitution
of the linear terms in the bra and ket wave functions of mixing type
in Eq.~(\ref{eq:Psi1sc}) gives two terms for this contribution as

\begin{equation}
\langle f|c_{1\uparrow}|0\rangle_{\rm cs}^{1}=\langle\Psi_{cs,f}^{1}|a_{1}D_{1}S_{1}^{-}|\Psi_{0}^{0}\rangle+\langle\Psi_{f}^{0}|a_{1}D_{1}S_{1}^{-}|\Psi_{cs,0}^{1}\rangle.\label{eq:ME_sc1def}
\end{equation}
We will deal with the first term in the r.h.s. of the above expression
first. Since $\left|\Psi_{\rm cs}^{1}\right\rangle $ in Eq.~(\ref{eq:Psi1sc})
is linear in the charge coordinates and in the charge quasimomenta,
the sum over the charge coordinates can be evaluated independently
of the spin coordinates giving the ``average'' values of the charge
quasimomenta of the bra state that depend on the coordinate on the
spin chain,
\begin{align}
\bar{k}_{l}^{f0} & =\frac{1}{L^{N-1}\left(N-1\right)!}\sum_{j_{2},\dots,j_{N},Q,Q'}\left(-1\right)^{Q+Q'}e^{iQ\mathbf{k}_{0}^{0}\cdot O\left(1,j_{2},\dots,j_{N}\right)-iQ'\mathbf{k}_{f}^{0}\cdot O\left(j_{2},\dots,j_{N}\right)}Q'k_{l}^{f0},
\end{align}
which in the thermodynamic limit gives 
\begin{equation}
\bar{k}_{l}^{f0}=\langle f|c_{1\uparrow}|0\rangle_{\rm c}^{0}P_{f}\left[N^{-1}+i\cot\left(\frac{P_{\rm s}^{f}-P_{\rm s}^{0}}{2}\right)\left(\delta_{l,1}-\delta_{l,N-1}\right)\right].\label{eq:k_lf0bar}
\end{equation}

Substitution of the latter in the remaining sum over the spin coordinates
gives the first term in Eq.~(\ref{eq:ME_sc1def}) as 
\begin{multline}
\langle\Psi_{cs,f}^{1}|a_{1}D_{1}S_{1}^{-}|\Psi_{0}^{0}\rangle=\langle f|c_{1\uparrow}|0\rangle_{\rm c}^{0}\frac{4P_{f}}{Z_{f}Z_{0}}\sum_{R,R',x_{2},\dots,x_{M}}e^{i\sum_{l<m}\varphi_{R_{l}R_{m}}^{0}-i\sum_{l<m}\varphi_{R_{l}'R_{m}'}^{f}+iR\mathbf{q}_{0}^{0}\cdot\left(1,x_{2},\dots,x_{M}\right)-iR'\mathbf{q}_{0}^{f}\cdot\left(x_{2},\dots,x_{M}\right)}\\
\times i\sum_{m'}\left[\frac{1-e^{-iR'q_{m'}^{f0}}}{2N}+\frac{1-\cos R'q_{m'}^{f0}}{N}x_{m'}+i\cot\left(\frac{P_{\rm s}^{f}-P_{\rm s}^{0}}{2}\right)\left(1-\cos R'q_{m'}^{f0}\right)\right] ,\label{eq:ME_sc1_part_1_v2}
\end{multline}
where the $\delta$-functions originating from Eq.~(\ref{eq:k_lf0bar})
are resolved by the sums over $m'$ in Eq.~(\ref{eq:Psi1sc}) and
the thermodynamic limit is also taken. The first and third terms
in the second line of the last expression can be taken outside of
the sum over the permutations $R'$ and $R$ since under the sum over
$m'$ they do not depend on $R'$ or $R$, and, then, the remaining
sum over the spin coordinates for them in the first line of the above
expression is $\langle f|c_{1\uparrow}|0\rangle_{\rm s}^{0}$ that was
already evaluated in Eq.~(\ref{eq:ME_s0}). Together, the first and
third terms in Eq.~(\ref{eq:ME_sc1_part_1_v2}) give
\begin{equation}
\langle\Psi_{cs,f}^{1}|a_{1}D_{1}S_{1}^{-}|\Psi_{0}^{0}\rangle=\langle f|c_{1\uparrow}|0\rangle^{0}\sum_{m}\left[2i\frac{1-e^{-iq_{m}^{f0}}}{N}-4\cot\left(\frac{P_{\rm s}^{f}-P_{\rm s}^{0}}{2}\right)\left(1-\cos q_{m}^{f0}\right)\right].\label{eq:ME_sc1_part_1_v3}
\end{equation}
The second term in the second line in Eq.~(\ref{eq:ME_sc1_part_1_v2})
depends explicitly on the spin coordinates. However, this dependence
has the same form as the linear term of the bra wave function of the
spin type in Eq.~(\ref{eq:Psi1s}) with $\mathbf{q}_{f}^{1}=4P_{f}\left(\cos\mathbf{q}_{f}^{0}-\mathbf{1}\right)/N$
when we realise that the derivative of the two-magnon scattering phase
in Eq.~(\ref{eq:Psi1s}) is zero for $\mathbf{q}_{f}^{1}=4P_{f}\left(\cos\mathbf{q}_{f}^{0}-\mathbf{1}\right)/N$, 
since 
\begin{equation}
\frac{\left(1-\cos R'q_{l}^{f0}\right)\left(1-\cos R'q_{m}^{f0}\right)-\left(1-\cos R'q_{m}^{f0}\right)\left(1-\cos R'q_{m}^{f0}\right)}{\left(e^{iR'q_{l}^{f0}}+e^{-iR'q_{m}^{f0}}-2\right)\left(e^{-iR'q_{l}^{f0}}+e^{iR'q_{m}^{f0}}-2\right)}=0.
\end{equation}
Then, the sum over the spin coordinates in the second term in Eq.~(\ref{eq:ME_sc1_part_1_v2})
can be evaluated using the same trick as for the spin contribution
to the linear term in Eq.~(\ref{eq:ME_s1_part_2_def}), giving the
same result as in Eq.~(\ref{eq:ME_s1_part_2}) but with $\mathbf{q}_{f}^{1}=4P_{f}\left(\cos\mathbf{q}_{f}^{0}-\mathbf{1}\right)/N$.

Now we deal with the second term in the r.h.s.\ of Eq.~(\ref{eq:ME_sc1def}).
Similarly to the first term, it is linear in the charge coordinates
and in the quasimomenta. Therefore, the sum over the charge coordinates
can be evaluated independently of the spin coordinates, giving the ``average''
values of the charge quasimomenta of the ket state that depend on
the coordinate on the spin chain,
\begin{equation}
\bar{k}_{l}^{00}=\frac{1}{L^{N-1}\left(N-1\right)!}\sum_{j_{2},\dots,j_{N},Q,Q'}\left(-1\right)^{Q+Q'}e^{iQ\mathbf{k}_{0}^{0}\cdot O\left(1,j_{2},\dots,j_{N}\right)-iQ'\mathbf{k}_{f}^{0}\cdot O\left(j_{2},\dots,j_{N}\right)}Qk_{l}^{00},
\end{equation}
which in the thermodynamic limit gives 
\begin{equation}
\bar{k}_{l}^{00}=\langle f|c_{1\uparrow}|0\rangle_{\rm c}^{0}P_{f}\left[N^{-1}-\delta_{l,1}+i\cot\left(\frac{P_{\rm s}^{f}-P_{\rm s}^{0}}{2}\right)\left(\delta_{l,2}-\delta_{l,N}\right)\right].\label{eq:k_l00bar}
\end{equation}

Substitution of the latter into the remaining sum over the spin coordinates
gives the second term in Eq.~(\ref{eq:ME_sc1def}) as 
\begin{multline}
\langle\Psi_{f}^{0}|a_{1}D_{1}S_{1}^{-}|\Psi_{cs,0}^{1}\rangle=\langle f|c_{1\uparrow}|0\rangle_{\rm c}^{0}\frac{4P_{f}}{Z_{f}Z_{0}}\sum_{R,R',x_{2},\dots,x_{M}}e^{i\sum_{l<m}\varphi_{R_{l}R_{m}}^{0}-i\sum_{l<m}\varphi_{R_{l}'R_{m}'}^{f}+iR\mathbf{q}_{0}^{0}\cdot\left(1,x_{2},\dots,x_{M}\right)-iR'\mathbf{q}_{0}^{f}\cdot\left(x_{2},\dots,x_{M}\right)}\\
\times\left(-i\right)\sum_{m'}\Bigg[\frac{1-e^{iRq_{m'}^{00}}}{2N}-\frac{1-e^{iRq_{m'}^{00}}}{2}\delta_{x_{m'},1}+\frac{1-\cos Rq_{m'}^{00}}{N}x_{m'}+\left(1-\cos Rq_{m'}^{00}\right)\\
+i\left(1-\cos Rq_{m'}^{00}\right)\cot\left(\frac{P_{\rm s}^{f}-P_{\rm s}^{0}}{2}\right)-i\left(1-\cos Rq_{m'}^{00}\right)\cot\left(\frac{P_{\rm s}^{f}-P_{\rm s}^{0}}{2}\right)\delta_{x_{m'},1}\Bigg],\label{eq:ME_sc1_part_2_v2}
\end{multline}
where the $\delta$-functions originating from Eq.~(\ref{eq:k_l00bar})
are resolved by the sums in Eq.~(\ref{eq:Psi1sc}) and the thermodynamic
limit is also taken. Similarly to the contributions in Eq.~(\ref{eq:ME_sc1_part_1_v3}),
the first and the fourth terms in the second line and the first term
in the third line of the above expression are independent of the spin
coordinates and the sum over the permutation $R$ and $R'$, so they
can be taken outside of them, and the remaining sums give $\langle f|c_{1\uparrow}|0\rangle_{\rm s}^{0}$, which was already evaluated in Eq.~(\ref{eq:ME_s0}). The third term
in the second line in Eq.~(\ref{eq:ME_sc1_part_2_v2}) has the same
dependence on the spin coordinates as the linear term of the ket wave
function of the spin type in Eq.~(\ref{eq:Psi1s}) and can be evaluated
in the same way as the corresponding term in Eq.~(\ref{eq:ME_sc1_part_1_v2}),
giving the spin contribution to the linear term in Eq.~(\ref{eq:ME_s1_part_1})
but with $\mathbf{q}_{0}^{1}=4P_{f}\left(\cos\mathbf{q}_{0}^{0}-\mathbf{1}\right)/N$.

The remaining two terms with $\delta_{x_{m}',1}$ in Eq.~(\ref{eq:ME_sc1_part_2_v2})
do not have analogs in the first term in Eq.~(\ref{eq:ME_sc1_part_1_v2}).
The second term in the second line in Eq.~(\ref{eq:ME_sc1_part_2_v2})
has the first part that is independent of $Rq_{m'}^{00}$ and gives
$\langle f|c_{1\uparrow}|0\rangle_{\rm s}^{0}$ as 
\begin{equation}
\frac{1}{2}\frac{1}{Z_{f}Z_{0}}\sum_{R,R'x_{2},\dots,x_{M}}e^{i\sum_{l<m}\varphi_{R_{l}R_{m}}^{0}-i\sum_{l<m}\varphi_{R_{l}'R_{m}'}^{f}+iR\mathbf{q}_{0}^{0}\cdot\left(1,x_{2},\dots,x_{M}\right)-iR'\mathbf{q}_{0}^{f}\cdot\left(x_{2},\dots,x_{M}\right)}=\frac{1}{2}\langle f|c_{1\uparrow}|0\rangle_{\rm s}^{0}.
\end{equation}
The other contribution to this term with $Rq_{m'}^{00}$ can be represented
in the thermodynamic limit as a shift of the spin chain to the right
by one site, evaluated as a spin part of the matrix element in the
zeroth order, and reverse shifted back by one site, giving
\begin{multline}
-\frac{1}{2}\frac{1}{Z_{f}Z_{0}}\sum_{R,x_{2},\dots,x_{M}}e^{i\sum_{l<m}\varphi_{R_{l}R_{m}}^{0}-i\sum_{l<m}\varphi_{R_{l}'R_{m}'}^{f}+iR\mathbf{q}_{0}^{0}\cdot\left(1,x_{2},\dots,x_{M}\right)-iR'\mathbf{q}_{0}^{f}\cdot\left(x_{2},\dots,x_{M}\right)}e^{iRq_{1}^{00}}\\
=-\frac{1}{2}e^{-i\left(P_{\rm s}^{0}-P_{\rm s}^{f}\right)}\langle f|c_{1\uparrow}|0\rangle_{\rm s}^{0}.\label{eq:ME_sc1_part_2_spin_shift_1}
\end{multline}

The second term in the third line in Eq.~(\ref{eq:ME_sc1_part_2_v2})
has the first part that is independent of $Rq_{m'}^{00}$ and gives
$\langle f|c_{1\uparrow}|0\rangle_{\rm s}^{0}$ as 
\begin{multline}
-i\cot\left(\frac{P_{\rm s}^{f}-P_{\rm s}^{0}}{2}\right)\frac{1}{Z_{f}Z_{0}}\sum_{R,R',x_{2},\dots,x_{M}}e^{i\sum_{l<m}\varphi_{R_{l}R_{m}}^{0}-i\sum_{l<m}\varphi_{R_{l}'R_{m}'}^{f}+iR\mathbf{q}_{0}^{0}\cdot\left(1,x_{2},\dots,x_{M}\right)-iR'\mathbf{q}_{0}^{f}\cdot\left(x_{2},\dots,x_{M}\right)}\\
=-i\cot\left(\frac{P_{\rm s}^{f}-P_{\rm s}^{0}}{2}\right)\langle f|c_{1\uparrow}|0\rangle_{\rm s}^{0}.
\end{multline}
In the other contribution $\cos Rq_{m'}^{00}$ can be represented
as a sum of two exponential functions,
\begin{equation}
\cos Rq_{m'}^{00}=\frac{e^{iRq_{m'}^{00}}+e^{-iRq_{m'}^{00}}}{2},
\end{equation}
and the same trick with the back-and-forth shift of the spin chain
as in Eq.~(\ref{eq:ME_sc1_part_2_spin_shift_1}) applied to each
of the two terms above gives
\begin{multline}
i\cot\left(\frac{P_{\rm s}^{f}-P_{\rm s}^{0}}{2}\right)\frac{1}{Z_{f}Z_{0}}\sum_{R,R'x_{2},\dots,x_{M}}e^{i\sum_{l<m}\varphi_{R_{l}R_{m}}^{0}-i\sum_{l<m}\varphi_{R_{l}'R_{m}'}^{f}+iR\mathbf{q}_{0}^{0}\cdot\left(1,x_{2},\dots,x_{M}\right)-iR'\mathbf{q}_{0}^{f}\cdot\left(x_{2},\dots,x_{M}\right)}\\
\times\sum_{m'}\cos Rq_{m'}^{00}\delta_{x_{m'},1}=i\cot\left(\frac{P_{\rm s}^{f}-P_{\rm s}^{0}}{2}\right)\cos\left(P_{\rm s}^{0}-P_{\rm s}^{f}\right)\langle f|c_{1\uparrow}|0\rangle_{\rm s}^{0}.
\end{multline}
Together, all the terms without $x_{m'}$ in Eq.~(\ref{eq:ME_sc1_part_2_v2})
give
\begin{multline}
\langle\Psi_{f}^{0}|a_{1}D_{1}S_{1}^{-}|\Psi_{cs,0}^{1}\rangle=\langle f|c_{1\uparrow}|0\rangle_{\rm s}^{0}P_{f}\Bigg[\sum_{m}\Bigg(-2i\frac{1-e^{iq_{m'}^{00}}}{N}+4\left(1-\cos q_{m'}^{00}\right)\cot\left(\frac{P_{\rm s}^{f}-P_{\rm s}^{0}}{2}\right)\\
-4i\left(1-\cos q_{m}^{00}\right)\Bigg)-2i\left(1-e^{-i\left(P_{\rm s}^{0}-P_{\rm s}^{f}\right)}\right)+4\cot\left(\frac{P_{\rm s}^{f}-P_{\rm s}^{0}}{2}\right)\left(\cos\left(P_{\rm s}^{0}-P_{\rm s}^{f}\right)-1\right)\Bigg]\label{eq:ME_sc1_part_2_v3}
\end{multline}
and the term with $x_{m'}$ gives the result in Eq.~(\ref{eq:ME_s1_part_1})
with $\mathbf{q}_{0}^{1}=4P_{f}\left(\cos\mathbf{q}_{0}^{0}-\mathbf{1}\right)/N$.

Under substitution of both contributions in Eqs.~(\ref{eq:ME_sc1_part_1_v3})
and (\ref{eq:ME_sc1_part_2_v3}) into the spin-charge mixing term in
 linear order in Eq.~(\ref{eq:ME_sc1def}), the whole matrix element
in the zeroth order factorises as 
\begin{equation}
\langle f|c_{1\uparrow}|0\rangle_{\rm cs}^{1}=\langle f|c_{1\uparrow}|0\rangle^{0}T_{\rm cs},
\end{equation}
where
\begin{equation}
T_{\rm cs}=P_{f}\left[4\frac{\sum_{m}\cos q_{m}^{f0}-\sum_{m}\cos q_{m}^{00}+\cos\left(P_{\rm s}^{0}-P_{\rm s}^{f}\right)}{\tan\left(\frac{P_{\rm s}^{f}-P_{\rm s}^{0}}{2}\right)}+2i\frac{\sum_{m}e^{iq_{m}^{00}}-\sum_{m}e^{-iq_{m}^{f0}}}{N}+2ie^{i\left(P_{\rm s}^{f}-P_{\rm s}^{0}\right)}\right].\label{eq:T_cs}
\end{equation}
The purely imaginary contributions were omitted since they do not contribute
to the linear order of the modulus squared of this matrix element
$|\langle f|c_{1\uparrow}|0\rangle|^{2}$ needed for observables in
this work, and the terms were rearranged for compactness. The contributions
with $x_{m'}$ are added by shifting the spin quasimomenta in the
spin contribution to the linear order in Eq.~(\ref{eq:Ts}) as 
\begin{equation}
\mathbf{q}_{0/f}^{1}\rightarrow\mathbf{q}_{0/f}^{1}+4P_{f}\frac{\cos\mathbf{q}_{0/f}^{0}-\mathbf{1}}{N}.
\end{equation}
The result in Eq.~(\ref{eq:T_cs}) is presented in Eq.~(9) of the
main text.

The matrix element of the creation operators $c_{k\uparrow}^{\dagger}$,
$\langle f|c_{1\uparrow}^{\dagger}|0\rangle$, can be expanded in
the same way in the Taylor series in $t/U$. In the zeroth order, the
expressions are the same as in Eqs.~(\ref{eq:MEminus}--\ref{eq:R_ab00})
and in the linear order the expressions are the same as in Eqs.~(\ref{eq:T_c}, \ref{eq:Ts}, \ref{eq:T_cs}),
in which, in both orders, the quasimomenta are swapped as $\mathbf{k}^{f},\mathbf{q}^{f}\leftrightarrow\mathbf{k}^{0},\mathbf{q}^{0}$
and the numbers of particles and spins are increased by one, $N\rightarrow N+1$
and $M\rightarrow M+1$.

\section{$t/U$ correction to normalisation factor}

We have selected the normalisation factor for the Lieb-Wu wave function
in Eq.~(\ref{eq:psi_jalpha}) to be unity in the $U=\infty$ limit,
$\langle\Psi^{0}|\Psi^{0}\rangle=1$, and have used it to evaluate
the linear order of the $t/U$ expansion of the matrix element in
the previous section. However, in the $t/U$ expansion calculated
in this work, deviations of the normalisation factor from unity also
have to be accounted for upto the same linear order, 
\begin{equation}
\langle\Psi|\Psi\rangle=1+\delta Z\frac{t}{U}.
\end{equation}
Under substitution of the $t/U$ expansion of the wave function up to
the linear order in Eq.~(\ref{eq:Psi01}) in the l.h.s.\ of the above
equation, we obtain the linear coefficient $\delta Z$ in its r.h.s.\ as 
\begin{equation}
\delta Z=\delta Z_{\rm c}+\delta Z_{\rm cs}+\delta Z_{\rm s},\label{eq:deltaZ_def}
\end{equation}
where the three contributions correspond to the three contributions
in the linear term of the wave function in Eq.~(\ref{eq:Psi1}).

The first term in Eq.~(\ref{eq:deltaZ_def}), the charge part, has
two contributions originating from the linear terms of the bra and
ket wave functions of the charge type in Eq.~(\ref{eq:Psi1c}), similarly
to the charge part of the matrix element in Eq.~(\ref{eq:ME_c1def}),
\begin{equation}
\delta Z_{\rm c}=\frac{1}{L^{N}N!}\sum_{Q,Q',\mathbf{j}}\left(-1\right)^{Q+Q'}e^{i\left(Q\mathbf{k}^{0}-Q'\mathbf{k}^{0}\right)\cdot\mathbf{j}}i\left(Q\mathbf{k}^{1}-Q'\mathbf{k}^{1}\right)\cdot\mathbf{j}.\label{eq:deltaZ_c_def}
\end{equation}
The scalar product in the last factor in the summand in the above
expression is a sum over $N$ terms with the coordinate of only one
particle in each term. The sum over the remaining $N-1$ coordinates
in each term produces a product of $N-1$ delta-functions in the charge
quasimomenta as $\delta\left(Qk_{j}-Q'k_{j}\right)$ so that the $N^{{\rm th}}$
charge quasimomenta in both permutation $Q\mathbf{k}$ and $Q'\mathbf{k}$
have to coincide producing, altogether, the $\delta$-function in
the whole permutations, $\delta\left(Q-Q'\right)$, under the sum
over the charge coordinates $\mathbf{j}$. Then, the sum over $Q'$
resolves the $\delta\left(Q-Q'\right)$ making the last factor in
the summand and the whole sum over the charge coordinates $\mathbf{j}$
and the permutations $Q$ and $Q'$ in Eq.~(\ref{eq:deltaZ_c_def})
zero, 
\begin{equation}
\delta Z_{\rm c}=0.\label{eq:deltaZ_c}
\end{equation}

The third term in Eq.~(\ref{eq:deltaZ_def}), the spin part, has
two complex-conjugated contributions also originating from the linear
terms in the bra and ket wave functions of the spin type in Eq.~(\ref{eq:Psi1s}),
\begin{equation}
\delta Z_{\rm s}=2{\rm Re}B_{\rm s},\label{eq:deltaZ_s_Def}
\end{equation}
where one of the contributions is 
\begin{multline}
B_{\rm s}=\frac{1}{Z^{2}}\sum_{R,R',\mathbf{x}}e^{i\sum_{l<m}\varphi_{R_{l}R_{m}}-i\sum_{l<m}\varphi_{R_{l}'R_{m}'}+i\left(R\mathbf{q}^{0}-R'\mathbf{q}^{0}\right)\cdot\mathbf{x}}\\
\left[i2\sum_{l<m}\frac{Rq_{l}^{1}\left(1-\cos Rq_{m}^{0}\right)-Rq_{m}^{1}\left(1-\cos Rq_{m}^{0}\right)}{\left(e^{iRq_{l}^{0}}+e^{-iRq_{m}^{0}}-2\right)\left(e^{-iRq_{l}^{0}}+e^{iRq_{m}^{0}}-2\right)}+iR\mathbf{q}^{1}\cdot\mathbf{x}\right].
\end{multline}
The sum over the spin coordinates in the last expression can be expressed
through the scalar product of two Bethe states using the trick with
the shift by $g\mathbf{q}^{1}$ and the limit of a derivative in Eqs.~(\ref{eq:Psi_s1_dPsi_s0}),
similarly to the spin contribution to the linear term of the matrix
element in Eq~(\ref{eq:ME_s1_part_1_def}),
\begin{equation}
B_{\rm s}=\lim_{g\rightarrow0}d_{g}\left\langle \mathbf{q}^{0}|\mathbf{q}^{0}+g\mathbf{q}^{1}\right\rangle .\label{eq:Bs}
\end{equation}
The scalar product of two Bethe states in the last expression can
be evaluated using the algebraic Bethe-ansatz method, giving the
Slavnov formula in Eq.~(\ref{eq:Scalar_product_ABA}). Taking the
$\Delta\rightarrow1$ limit of this result, as in Eq.~(\ref{eq:Heisenberg_limit_ABA}),
and dividing it by the square root of the prefactor in front of the
determinant in Eq.~(\ref{eq:Z2_ABA_coordinate}) for the bra and
ket states to account for the conversion of the spin normalisation
factor from the algebraic to the coordinate representation, we obtain
\begin{equation}
\left\langle \mathbf{q}^{0}|\mathbf{q}^{0}+g\mathbf{q}^{1}\right\rangle =\frac{1}{Z^{2}}\frac{\prod_{l\neq m}^{M}\left(\cot\frac{q_{l}^{0}+gq_{l}^{1}}{2}-\cot\frac{q_{m}^{0}}{2}+2i\right)\det\hat{Q}}{\prod_{l\neq m}^{M}\sqrt{\cot\frac{q_{l}^{0}+gq_{l}^{1}}{2}-\cot\frac{q_{m}^{0}+gq_{m}^{1}}{2}+2i}\prod_{l\neq m}^{M}\sqrt{\cot\frac{q_{l}^{0}}{2}-\cot\frac{q_{m}^{0}}{2}+2i}},\label{eq:q0_q0+gq1}
\end{equation}
where the matrix elements of $\hat{Q}$ are 
\begin{equation}
Q_{ab}=\frac{i\left(\cot\frac{q_{a}^{0}+gq_{a}^{1}}{2}-\cot\frac{q_{a}^{0}}{2}+2i\right)\left(e^{iN\left(q_{b}^{0}+gq_{b}^{1}\right)}\prod_{l=1\neq a}^{M}\frac{\cot\frac{q_{b}^{0}+gq_{b}^{1}}{2}-\cot\frac{q_{l}^{0}}{2}-2i}{\cot\frac{q_{b}^{0}+gq_{b}^{1}}{2}-\cot\frac{q_{l}^{0}}{2}+2i}-1\right)}{2\sin\frac{q_{a}^{0}+gq_{a}^{1}}{2}\sin\frac{q_{a}^{0}}{2}\left(\cot\frac{q_{b}^{0}+gq_{b}^{1}}{2}-\cot\frac{q_{a}^{0}}{2}\right)\left(\cot\frac{q_{b}^{0}+gq_{b}^{1}}{2}-\cot\frac{q_{a}^{0}}{2}+2i\right)}.\label{eq:Q_ab}
\end{equation}

Repeating the steps as for the spin contribution to the linear term
of the matrix element, we expand the matrix $\hat{Q}$ in the last
expression in a Taylor series in $g$ up to the linear order 
\begin{equation}
\hat{Q}_{0}+g\hat{Q}_{1},\label{eq:Q0+gQ1}
\end{equation}
as in Eq.~(\ref{eq:R_00+gR_01}). Here, however, we need to evaluate
the expansion coefficients with a bit more care, as limits of $g\rightarrow0$,
since both numerator and denominator in the diagonal matrix elements
$Q_{aa}$ in Eq.~(\ref{eq:Q_ab}) become zero for $g=0$. We evaluate
the zeroth-order term in Eq.~(\ref{eq:Q0+gQ1}) as 
\begin{equation}
\hat{Q}_{0}=\lim_{g\rightarrow0}\hat{Q}
\end{equation}
and obtain the same matrix as in the Gaudin normalisation factor
\citep{Gaudin81} in Eq.~(\ref{eq:Q_ml0}). Calculating the coefficient
in the linear order as 
\begin{equation}
\hat{Q}_{1}=\lim_{g\rightarrow0}d_{g}\hat{Q},
\end{equation}
we obtain
\begin{equation}
Q_{aa}^{1}=\frac{q_{a}^{1}}{2}\left[\sum_{l\neq a}\frac{\sin^{2}\frac{q_{l}^{0}}{2}\left(2\sin q_{a}^{0}-\sin\left(q_{a}^{0}+q_{l}^{0}\right)\right)}{\left(\frac{3}{2}+\frac{\cos\left(q_{a}^{0}+q_{l}^{0}\right)}{2}-\cos q_{a}^{0}-\cos q_{l}^{0}\right)^{2}}+i\left(N-\sum_{l\neq a}\frac{1-\cos q_{l}^{0}}{\frac{3}{2}+\frac{\cos\left(q_{a}^{0}+q_{l}^{0}\right)}{2}-\cos q_{a}^{0}-\cos q_{l}^{0}}\right)^{2}\right],\label{eq:Q_aa1}
\end{equation}
\begin{multline}
Q_{ab}^{1}=i\frac{1-\cos q_{b}^{0}}{3+\cos\left(q_{a}^{0}+q_{b}^{0}\right)-2\cos q_{a}^{0}-2\cos q_{b}^{0}}\Bigg[q_{b}^{1}\Bigg(N-\frac{1}{2}-\sum_{l\neq a,b}\frac{1-\cos q_{l}^{0}}{\frac{3}{2}+\frac{\cos\left(q_{b}^{0}+q_{l}^{0}\right)}{2}-\cos q_{b}^{0}-\cos q_{l}^{0}}\\
-2\frac{e^{i\frac{q_{a}^{0}-3q_{b}^{0}}{2}}\sin\frac{q_{a}^{0}}{2}\left(1+e^{iq_{a}^{0}}+e^{iq_{b}^{0}}\left(e^{iq_{a}^{0}}-3\right)\right)}{\sin\frac{q_{b}^{0}}{2}\left(e^{iq_{a}^{0}}+e^{-iq_{b}^{0}}-2\right)^{2}}\Bigg)\frac{e^{iq_{a}^{0}+iq_{b}^{0}}+1-2e^{iq_{b}^{0}}}{e^{iq_{a}^{0}}-e^{iq_{b}^{0}}}+q_{a}^{1}\frac{1+i\sin q_{a}^{0}}{1-\cos q_{a}^{0}}\Bigg]\label{eq:Q_ab1}
\end{multline}
for the diagonal and off-diagonal $a\neq b$ matrix elements of
$\hat{Q}_{1}$ respectively. In the substitution of the expansion
in Eq.~(\ref{eq:Q0+gQ1}) into Eq.~(\ref{eq:Bs}) we use the Jacobi
formula for the derivative of a determinant as in Eq.~(\ref{eq:d_gdetC0+gC1}).
Then we substitute the result for $B_{\rm s}$ into Eq.~(\ref{eq:deltaZ_s_Def})
and obtain
\begin{equation}
\delta Z_{\rm s}=2{\rm Re}{\rm Tr}\left(\hat{Q}_{0}^{-1}\hat{Q}_{1}\right),\label{eq:deltaZ_s}
\end{equation}
where the linear term in $g$ coming from the Taylor expansion of
the prefactor in front of the determinant in Eq.~(\ref{eq:q0_q0+gq1})
does not contribute since it has zero real part.

The second term in Eq.~(\ref{eq:deltaZ_def}), in which the sums
over the spin and charge coordinates are mixed with each other, has
two complex-conjugated contributions also originating from the linear
terms in the bra and ket wave functions of the spin-charge mixing
type in Eq.~(\ref{eq:Psi1sc}),
\begin{equation}
\delta Z_{sc}=2{\rm Re}B_{sc}.\label{eq:deltaZ_sc_def}
\end{equation}
Since $\left|\Psi_{\rm cs}^{1}\right\rangle $ in Eq.~(\ref{eq:Psi1sc})
is linear in the charge coordinates and in the charge quasimomenta,
the sum over the charge coordinates in $B_{sc}$ can be evaluated
independently of the spin coordinates, giving the ``average'' value of
the charge quasimomenta over the charge state that depend on the coordinate
on the spin chain,
\begin{equation}
\bar{k}_{l}^{0}=\frac{1}{L^{N}N!}\sum_{\mathbf{j},Q,Q'}\left(-1\right)^{Q+Q'}e^{i\left(Q\mathbf{k}^{0}-Q'\mathbf{k}^{0}\right)\cdot O\mathbf{j}}Qk_{l}^{0},
\end{equation}
which in the thermodynamic limit gives 
\begin{equation}
\bar{k}_{l}^{0}=\frac{P}{N}+i\frac{N}{2L}\left(\delta_{l,1}-\delta_{l,N}\right).\label{eq:k_l0bar}
\end{equation}

Substitution of the latter into the remaining sum over the spin coordinates
gives 
\begin{equation}
B_{sc}=-\frac{4i}{Z^{2}}\sum_{R,R',\mathbf{x}}e^{i\sum_{l<m}\varphi_{R_{l}R_{m}}-i\sum_{l<m}\varphi_{R_{l}'R_{m}'}+i\left(R\mathbf{q}_{0}-R'\mathbf{q}_{0}\right)\cdot\mathbf{x}}i\sum_{m'}\frac{N}{2L}\left(1-\cos Rq_{m'}^{0}\right)
\end{equation}
where $\delta$-functions originating from Eq.~(\ref{eq:k_l0bar})
are resolved by the sums in Eq.~(\ref{eq:Psi1sc}) and the thermodynamic
limit is also taken, similarly to the spin-charge mixing part in linear
order of the matrix element in Eq.~(\ref{eq:ME_sc1_part_1_v2}). The
sum over $m'$ in the last expression can be taken outside the
sum over the permutations $R$, since under the sum over $m'$ it does not
depend on $R$. Then, the remaining sum over the spin coordinates
is the normalisation factor $Z^{2}$ that cancels the $Z^{2}$ in
the denominator, and we obtain 
\begin{equation}
B_{sc}=-\frac{2N}{L}\sum_{m}\left(\cos q_{m}^{0}-1\right).
\end{equation}
Substitution of this expression back into Eq.~(\ref{eq:deltaZ_sc_def})
gives the spin-charge mixing contribution as 
\begin{equation}
\delta Z_{sc}=-\frac{4N}{L}\sum_{m}\left(\cos q_{m}^{0}-1\right).\label{eq:deltaZ_sc}
\end{equation}

Substitution of the three results in Eqs.~(\ref{eq:deltaZ_c}, \ref{eq:deltaZ_s}, \ref{eq:deltaZ_sc})
into Eq.~(\ref{eq:deltaZ_def}) gives the linear order of the $t/U$
expansion of the normalisation factor as 
\begin{equation}
\delta Z=2{\rm Re}{\rm Tr}\big(\hat{Q}_{0}^{-1}\hat{Q}_{1}\big)-\frac{4N}{L}\sum_{m}\big(\cos q_{m}^{0}-1\big).\label{eq:deltaZ}
\end{equation}
The result in Eq.~(\ref{eq:deltaZ}) is presented in Eqs.~(10) of the main paper. 

\section{Transport theory for the free system}
In order to assess expectations for the conductance measured in the transport-spectroscopy experiment in the nonlinear regime, we present here the corresponding transport theory for the noninteracting 1D system \cite{vianez_book}. Then we compare its prediction for the nonlinear charge mode in the particle sector with the data measured for our wires with Coulomb interactions.

The current between the two wells in the weak-tunneling regime for the device in the inset in Fig.~3A of the main paper is given by the convolution of the two spectral functions as \cite{Altland99}
\begin{equation}
I\left(B,V_{{\rm dc}}\right)=\int {\rm d}^{2}k\,{\rm d}E\left(f_{T}^{\rm UW}\left(E-eV_{{\rm dc}}\right)-f_{T}^{\rm LW}\left(E\right)\right)
A_{\rm UW}\left(\mathbf{k},E\right)A_{\rm LW}\left(\mathbf{k}+\frac{ed}{\hbar}\left(\mathbf{n}\times\mathbf{B}\right),E-eV_{{\rm dc}}\right),\label{eq:I_nonint}
\end{equation}
where $A_{\rm UW/LW} (\mathbf{k},E)$ and  $f_{T}^{\rm UW/LW}\left(E\right)$ are the spectral and Fermi functions for the upper/lower wells (UW/LW), $-e$ is the electron charge, $d$ is the distance between the centers of the wells, and $\mathbf{n}=\hat{\mathbf{z}}$ is the normal to the 2D plane of the wells. When a DC bias $V_{\rm dc}$ is applied between the wells, the energy offset acquired between the two electronic systems is $eV_{\rm dc}$ and an in-plane magnetic field $\mathbf{B}=-B\hat{\mathbf{y}}$ shifts the momentum $k$ in the tunneling process due to the Lorentz force by $edB$ in the $x$-direction.

The 2D electrons in the lower well are described by the spectral function of the Fermi gas as 
\begin{equation}
A_{\rm LW}\left(\mathbf{k},E\right)=\frac{1}{\pi}\frac{\Gamma}{\Gamma^{2}+\left(E-\frac{\hbar^{2}\left(k-k_{{\rm F}}^{{\rm 2D}}\right)}{2m_{{\rm 2D}}^{*}}^{2}\right)^{2}} ,\label{eq:A2D}
\end{equation}
where $k_{\rm F}^{\rm 2D}$ is its Fermi momentum, $m_{\rm 2D}^{*}=0.062 m_e$ \cite{Tan05} is the Fermi-liquid's effective electron mass, $m_e$ is the free electron mass, and $\Gamma$ is the width of the inhomogeneous broadening which we assume to be larger than the interaction broadening. The 1D electrons in the upper well are described by the spectral function of a Fermi system without interactions,
\begin{equation}
A_{\rm UW}\left(\mathbf{k},E\right)=\frac{1}{\pi}\frac{\Gamma}{\Gamma^{2}+\left(E-\frac{\hbar^{2}\left(k_{x}-k_{{\rm F}}^{1D}\right)}{2m_{0}}^{2}\right)^2},\label{eq:A1D_nonint}
\end{equation}
with the same inhomogeneous broadening $\Gamma$ and a free mass $m_0$.

\begin{figure}
\centering
\includegraphics[width=1\columnwidth]{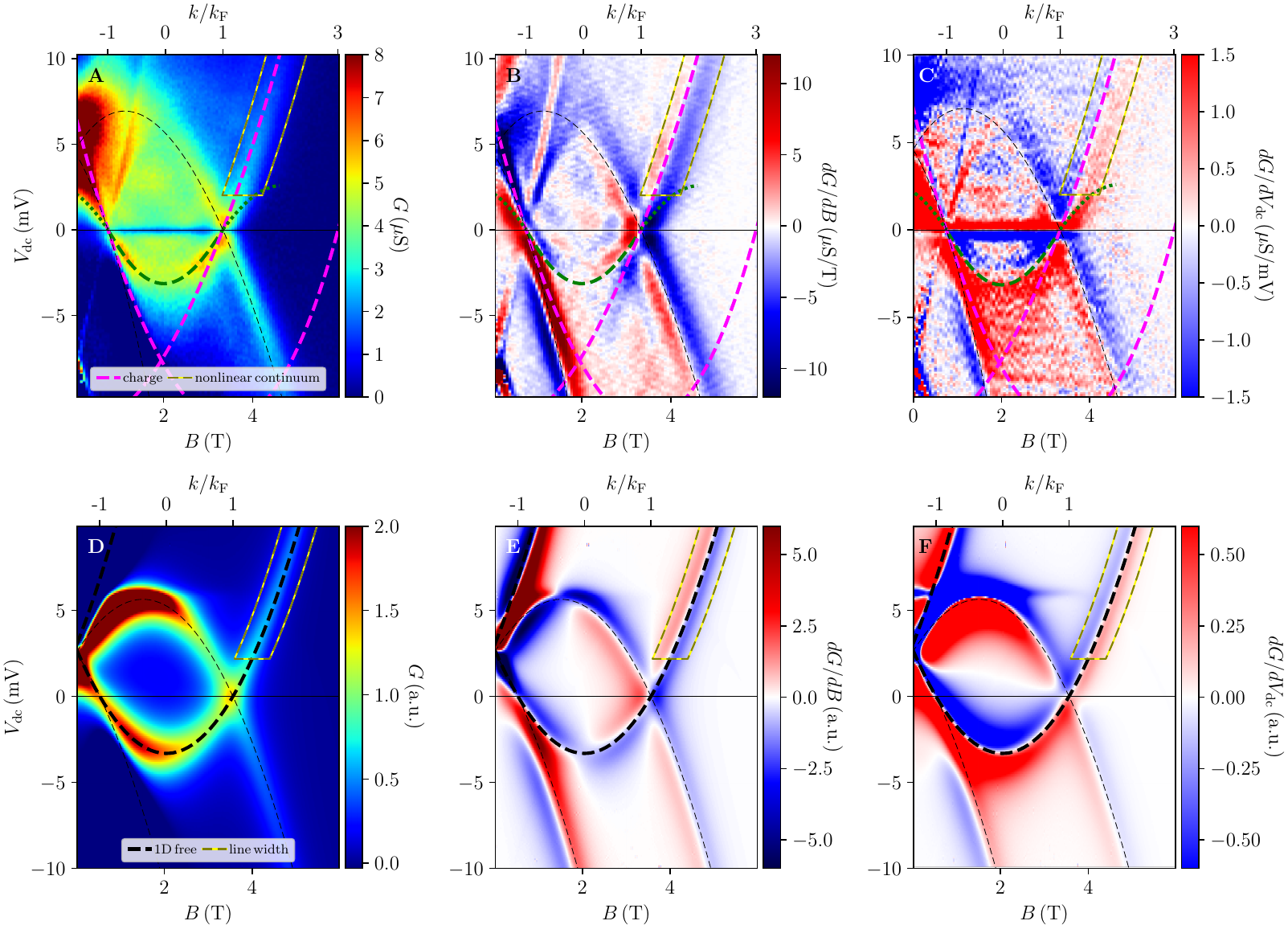}\caption{\textbf{A} The conductance $G\left(B,V_{\rm dc}\right)$ measured for $V_{\rm FG}=-664\, {\rm mV}$. The lines are the same as in Fig.\ 3A of the main paper. \textbf{B} and \textbf{C} are the ${\rm d}G/{\rm d}B$ and ${\rm d}G/{\rm d}V_{\rm dc}$ derivatives of the conductance in \textbf{A}. \textbf{D}  Numerical evaluation of the conductance for the transport theory for the non-interacting 1D system in Eqs.\ (\ref{eq:I_nonint}, \ref{eq:A2D},\ref{eq:A1D_nonint})  using $m_0=0.93 m_{\rm e}$, $\Gamma=0.3\,{\rm meV}$, $k_{\rm F}^{\rm 1D}=73.6\,\mu$m$^{-1}$, and  $k_{\rm F}^{\rm 2D}=99.7\,\mu$m$^{-1}$. The additional thick black dashed line is the 1D dispersion in Eq.\ (\ref{eq:A1D_nonint}). \textbf{E} and \textbf{F} are the ${\rm d}G/{\rm d}B$ and ${\rm d}G/{\rm d}V_{\rm dc}$ derivatives of the conductance in \textbf{D}.}\label{fig:SI_charge_particle_continuum}
\end{figure}

The numerical evaluation of the integrals in Eq.~(\ref{eq:I_nonint}) with the spectral functions in Eqs.~(\ref{eq:A2D}, \ref{eq:A1D_nonint}) for the parameters parameters comparable with our semiconductor experiment ($m_0=0.93 m_e$, $\Gamma=0.3\,{\rm meV}$, $k_{\rm F}^{\rm 1D}=73.6\, \mu m^{-1}$, and  $k_{\rm F}^{\rm 2D}=99.7\,\mu$m$^{-1}$)  is presented in Figs.~\ref{fig:SI_charge_particle_continuum}D-F as the differential conductance $G={\rm d}I/{\rm d}V_{\rm dc}$ and its derivatives ${\rm d}G/{\rm d}B$ and ${\rm d}G/{\rm d}V_{\rm dc}$. We observe that a symmetrical line width in the 1D spectral function without interaction in Eq.~(\ref{eq:A1D_nonint}) does not lead to any perceptible asymmetry of the 1D line in $G$ in the charge sector enclosed by the olive-yellow dashed line. On the other hand, the measured conductance $G$ for electrons with the Coulomb interaction, see Fig.~\ref{fig:SI_charge_particle_continuum}A-C, shows a large asymmetry of this line, which is a manifestation of the large continuum of nonlinear many-body excitations predicted in Fig.~2A of the main paper around this line in the particle sector.

\begin{figure}
\centering
\includegraphics[width=1\columnwidth]{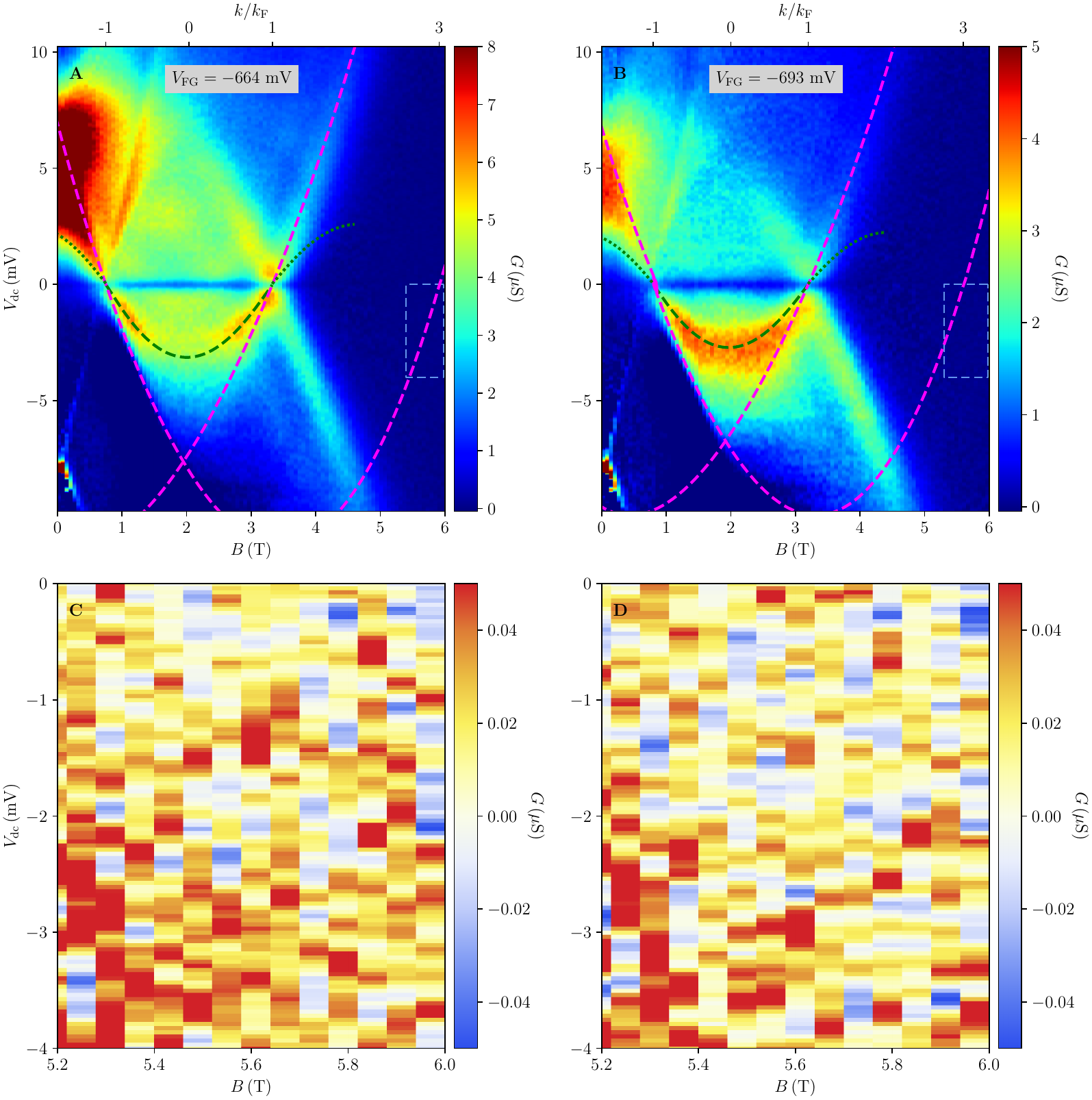}\caption{\textbf{A} and \textbf{B} are the conductance $G\left(B,V_{\rm dc}\right)$ measured for $V_{\rm FG}=-664\, {\rm mV}$ and $V_{\rm FG}=-693\, {\rm mV}$.  The green and magenta dashed lines are the dispersions of the spin and charge modes obtained from the full Lieb-Wu equations for $\gamma=1.25$ in \textbf{A}  and for $\gamma=1.30$ in \textbf{B}, and  were corrected for  capacitance using $c_{\rm UL}=5.6\,{\rm mFm^{-2}}$ and $c_{\rm UW}=4.7\,{\rm mFm^{-2}}$, see details in \cite{Vianez21}, for both applied $V_{\rm FG}$. The upper horizontal axis is the linear transformation of $B$ using the two crossing points with the $V_{\rm dc}=0$ line as $B_{\rm lo}=0.75\, {\rm T}$ is $-k_{\rm F}$ and $B_{\rm hi}=3.33\, {\rm T}$ is $k_{\rm F}$ in \textbf{A} and $B_{\rm lo}=0.80\, {\rm T}$ is $-k_{\rm F}$ and $B_{\rm hi}=3.20\, {\rm T}$ is $k_{\rm F}$ in \textbf{B}. \textbf{C} and \textbf{D} are the zoomed-in conductances in the light-blue dashed rectangle in \textbf{A} and \textbf{B} respectively. The noise floor in both measurements is $\left|G\right|\sim 0.05\, {\rm \mu S}$.}\label{fig:SI_3kF}
\end{figure}

\bibliographystyle{apsrev4-2}
\bibliography{citations}